\documentclass[aps,prx,twocolumn,showpacs,notitlepage,superscriptaddress]{revtex4-2}
\usepackage{graphicx}
\usepackage{color}\usepackage{enumerate}
\usepackage{amsmath}\usepackage{amssymb}\usepackage{stmaryrd}
\usepackage{multirow}
\usepackage{float}
\usepackage{longtable,booktabs}
\usepackage[caption=false]{subfig}
\usepackage{dsfont}
\usepackage{amssymb}
\usepackage{placeins}
\usepackage{mathtools}
\usepackage[pdftex,colorlinks=true,linkcolor=blue,citecolor=blue]{hyperref}

\usepackage{soul}
\def\half{\frac{1}{2}}

\DeclareMathAlphabet\mathbfcal{OMS}{cmsy}{b}{n}
\DeclarePairedDelimiter\bra{\langle}{\rvert}
\DeclarePairedDelimiter\ket{\lvert}{\rangle}
\DeclarePairedDelimiterX\braket[2]{\langle}{\rangle}{#1 \delimsize\vert #2}

\begin{document}
\title{
Spin-momentum locking from topological quantum chemistry:  applications to multifold fermions
}
\author{Mao Lin}
\affiliation{Department of Physics and Institute for Condensed Matter Theory, University of Illinois at Urbana-Champaign, Urbana, IL, 61801-3080, USA}%
\author{I\~{n}igo Robredo}
\affiliation{Donostia International Physics Center, 20018 Donostia-San Sebastian, Spain}
\affiliation{Max Planck Institute for Chemical Physics of Solids, 01187 Dresden, Germany}
\author{Niels B.~M. Schr\"{o}ter}
\affiliation{Max Planck Institute of Microstructure Physics, Weinberg 2, 06120 Halle, Germany}
\author{Claudia Felser}
\affiliation{Max Planck Institute for Chemical Physics of Solids, 01187 Dresden, Germany}
\author{Maia G. Vergniory}
\affiliation{Max Planck Institute for Chemical Physics of Solids, 01187 Dresden, Germany}
\affiliation{Donostia International Physics Center, 20018 Donostia-San Sebastian, Spain}
\author{Barry Bradlyn}%
\affiliation{Department of Physics and Institute for Condensed Matter Theory, University of Illinois at Urbana-Champaign, Urbana, IL, 61801-3080, USA}%
\date{\today}
\begin{abstract}
In spin-orbit coupled crystals, symmetries can protect multifold degeneracies with large Chern numbers and Brillouin zone spanning topological surface states. 
In this work, we explore the extent to which the nontrivial topology of chiral multifold fermions impacts the spin texture of bulk states.  
To do so, we formulate a definition of spin-momentum locking in terms of reduced density matrices. 
Using tools from the theory of topological quantum chemistry, we show how the reduced density matrix can be determined from the knowledge of the basis orbitals and band representation forming the multifold fermion. 
We show how on-site spin orbit coupling, crystal field splitting, and Wyckoff position multiplicity compete to determine the spin texture of states near chiral fermions. 
We compute the spin texture of multifold fermions in several representative examples from space groups $P432$ (207) and $P2_13$ (198). 
We show that the winding number of the spin around the Fermi surface can take many different integer values, from zero all the way to $\pm 7$.
Finally, we conclude by showing how to apply our theory to real materials using the example of PtGa in space group $P2_13$.
\end{abstract}
\maketitle

\tableofcontents

\section{Introduction}

Over the last several decades, the study of topological insulators and semimetals has highlighted the importance of spin-orbit coupling (SOC) in condensed matter systems. 
One of the relativistic corrections to the Schr\"{o}dinger equation, SOC is known to split the electronic energy levels in molecules and crystalline solids\cite{Kittel87}. 
In solid-state systems, SOC can split degeneracies in the band structure of materials, leading to the emergence of topological insulating gaps\cite{kane2005quantum,kane2005z,bernevig2006quantum,fu2007topologicala,Xia09} and topological Weyl\cite{wan2011topological,lv2015observation,xu2015discoverya}, Dirac\cite{liu2014stable,liu2014discovery,wang2012dirac}, nodal line\cite{Kim2015,Chan2016}, and multifold fermions\cite{Bradlyn2016,chang2018topological,chang2017unconventional}. 

In the absence of inversion symmetry, SOC leads to the splitting of the twofold degeneracy of energy bands away from 
time-reversal invariant points in the Brillouin zone. 
When inversion symmetry is broken at the boundary of a 3D system, or by a substrate for a thin-film, this is known as the Rashba effect; in simple two-band models the Rashba splitting leads to a characteristic locking of the electron spin $\mathbf{s}$ to its momentum $\mathbf{k}$, $\langle\mathbf{s}\rangle\propto \nabla V \times \mathbf{k}$, where $V$ is the inversion symmetry breaking potential. 
At the surfaces of 3D topological insulators (TIs)---where only one spin-nondegenerate band is present---spin momentum-locking has served as a smoking gun to identify topological surface states\cite{qi2011topological,hasan2010colloquium}. 
Experimentally, it was observed that the spin angular momentum forms a nontrivial texture surrounding the surface Dirac cone of the TI\cite{hsieh2009tunable}. 
Because the spin direction is locked with the direction of the momentum, backscattering between ${\bf k}$ and $-{\bf k}$ by nonmagnetic impurities is forbidden, which reflects the topological stability of the surface states\cite{raghu2010collective}. 
Relatedly, the surface states of topological semimetals (TSMs) have also been predicted to show spin textures\cite{wan2011topological,armitage2018weyl}. 
The spin textures of the Fermi arc states on surfaces of Weyl semimetal (WSM) has been observed in TaAs\cite{xu2016spin}. Crucially, however, these states do not feature strong spin-momentum locking.

SOC can also lead to the splitting of doubly degenerate bands in the bulk Brillouin zone of noncentrosymmetric 3D crystals (the so-called Dresselhaus effect). 
For chiral crystals--those noncentrosymmetric crystals with no orientation-reversing symmetries--this leads to the prediction that time-reversal invariant momenta (TRIMs) generically host topologically nontrivial fermions known as Kramers-Weyl fermions\cite{chang2018topological}. 
In simple models of Kramers-Weyl fermions the interplay of chirality and SOC leads to a radial locking between spin and momentum, whereas the spin texture in conventional WSMs is nonuniversal and does not display spin-momentum locking.\cite{gatti2020radial,kim2021kramers}.
These studies not only reveal the close connection between the spin-momentum locking and the transport properties of TSM, such as magnetoresistance, but also provide an avenue for electrically controllable spin texture for future spintronics applications\cite{sun2015helical,lin2020electric,he2019kramers}. 

In chiral crystals with additional crystal symmetries, it has recently been shown that in addition to Kramers-Weyl fermions there are also topologically nontrivial multifold fermions with three-, four-, or six-fold band degeneracies\cite{Bradlyn2016}. 
These fermions permit exotic chiral optical responses\cite{flicker2018chiral,chang2017unconventional}, and can be found in various magnetic materials\cite{cano2019multifold} even when time-reversal symmetry is broken. 
Chiral multifold fermions can have large Chern numbers and feature multiple Fermi arc surface stats that spanning large fractions of the Brillouin zone, as has recently been observed experimentally in B20 compounds such as PdGa\cite{schroter2020observation}, AlPt\cite{schroter2019chiral}, RhSi\cite{sanchez2019topological}, and CoSi\cite{rao2019observation}. 
Since multifold fermions arise due to the interplay of SOC and chiral crystal symmetries, it is natural to wonder whether they feature similar spin textures predicted for Kramers-Weyl fermions. 
Furthermore, the prevalence of materials with well-resolved multifold degeneracies near the Fermi level makes this a pressing question for experimentalists.
However, due to the large number of bands and complicated unit cells necessary to realize chiral multifold fermions, the nature of any spin-momentum locking in these systems has not been systematically explored. 

In order to remedy this, and to further explore the nature of spin-momentum locking in chiral topological semimetals, in this work we develop a general framework for determining the universal features of the spin texture of bands. 
To do so, we make use of the framework of topological quantum chemistry (TQC)\cite{NaturePaper}. 
TQC provides a mechanism to map position space data regarding the electronic orbitals in a material to momentum space information about the band structure.
One of the key concepts of TQC is the band representation\cite{Zak1981,Bacry1988}, which maps the spin-orbital basis of the electrons on sites in position space to the momentum space description of the electronic bands in the Brillouin zone (BZ). 
Band representations can be decomposed into elementary band representations (EBR), which are band representations themselves but cannot be
further subdivided in position space while preserving the symmetry operations of the system\cite{cano2018building}. 
Bands that do not transform as a (positive) linear combination of EBRs do not have a symmetric and localizable position-space description, and hence correspond to topologically nontrivial bands.
Relatedly, if an EBR with nodal points is fractionally filled with electrons, then the material will be a topological metal or semimetal\cite{po2017symmetry,cano2021band,po2020symmetry,song2018mapping}. 
With the EBR method, more than a hundred materials have been identified as candidate TSM materials\cite{vergniory2019high,wieder2021topological,bigmaterials-china,bigmaterials-ashvin,vergniory2021all}.

The position space inputs to TQC include not just the location and orbital character of the constituent electronic states of a material, but also the electron spin states.  
In particular, let us consider a set of atomic orbitals at a given point in the unit cell of a crystal. 
The orbitals at this point transform in a representation of the subgroup of the space group that leave the point invariant. 
This subgroup is known as the site-symmetry group. 
The site-symmetry group of a point can be viewed as the symmetry group of the local crystal field potential at that point; for spin-orbit coupled systems this local crystal field potential includes spin-orbit coupling, and so the representations of the site symmetry group are spanned by spin-orbit coupled basis states. 
For example, in an octahedral crystal the fundamental building blocks of electronic states are atomic $S_{1/2}$, $P_{1/2}$, $P_{3/2}$, and similar spin-orbit coupled states. 
These spin-orbit coupled basis states then determine the EBRs for the space group. 
TQC thus provides a mapping from local spin-orbit coupled orbitals to the Bloch states in the Brillouin zone, and so encodes information about the momentum space spin texture of band representations. 

In this work, we will show how to extract the spin texture near chiral multifold fermions using the technology of TQC. 
In particular, we will show how the bulk spin texture near chiral fermions is determined by the band representation and basis states forming the degeneracy. 
To do so, we will formulate a theory of reduced density matrices for spin states in Bloch bands, which takes into account the nontrivial entanglement between spin and orbital degrees of freedom implied by SOC and crystal field splitting.
Using this formalism we will show that spin-momentum locking in Kramers-Weyl and multifold fermions comes in a variety of forms, which can be nonquantized and vary depending on the number of band representations that are energetically close together. 
Our approach goes further than recent works in that we include the important effects of position space information in addition to momentum space $\mathbf{k}\cdot\mathbf{p}$ data, and generalizes previous results for Kramers-Weyl fermions\cite{chang2018topological,acosta2021different}.
Focusing on cubic crystals, we will derive the allowed patterns of spin-momentum locking for fourfold- and sixfold-degenerate multifold fermions at high-symmetry points. 
We will also see that spin-orbit entanglement in nonsymmorphic cubic crystals can lead to spin-orbit coupled (quadratic) twofold degeneracies with zero spin texture. 
Finally, we will show how our method can be used in conjunction with ab-initio calculations to study spin-momentum locking in realistic materials, with a focus on the newly-discovered multifold topological semimetal PtGa.

The structure of the paper is the following. 
In Sec.~\ref{sec:Spin reduced density matrices of irreducible band representations} we introduce the general theory of spin-reduced density matrices (sRDMs) and show how it applies to elementary band representations. 
In Sec.~\ref{sec:1apos}, as an example, we calculate the sRDM for the multiplicity-one Wyckoff positions in the chiral octahedral space group $P432$ (207). 
In Sec.~\ref{sec:Composite band representations} we describe the theory of sRDM for composite band representations and explore the interplay between on-site spin orbit coupling, crystal field splitting, and hybridization. 
To study the effect of band representations with more than one site per unit cell, in Sec.~\ref{sec:Band Representations from high-multiplicity Wyckoff Positions} we calculate the sRDM for multifold degeneracies in space group $P2_13$ (198) at the $\Gamma$ and $R$ points. 
We see that the action of the crystal symmetries on the sites within the unit cell leads to nontrivial entanglement between spin and orbital degrees of freedom, drastically changing the spin-momentum locking near band degeneracies. 
We then apply these results in Sec.~\ref{sec:abinit} to the experimentally relevant example of PtGa in space group $P2_13 $ (198), and compare our predicted spin-momentum locking with ab-intio density functional theory (DFT) calculations. 
In Sec.~\ref{sec:inpractice} we review the steps necessary to apply our method to other materials and space groups. 
We conclude in Sec.~\ref{sec:Discussion and Conclusion} with a discussion and outlook towards further extensions of our method.

\section{Spin reduced density matrices of irreducible band representations}
\label{sec:Spin reduced density matrices of irreducible band representations}

Given a set of bands with a chiral multifold degeneracy, we would like to identify how the real electron spin maps into the pseudospin space of the degeneracy. 
For degeneracies that occur in Wannier-representable bands, this can be done using the tools of topological quantum chemistry. 
In particular, any Wannier-representable set of bands can be described by a band representation of the space group. 
Upon choosing a basis for that band representation and restricting to the degeneracy of interest, we can obtain a mapping between position space and pseudospin space. 
Following the notation of Refs.~\cite{NaturePaper,cano2018building,cano2021band}, a band representation $\rho_\mathbf{k}=(\rho\uparrow G)_{\mathbf{q}}$ induced from a site-symmetry group representation $\rho$ at Wyckoff position $\mathbf{q}$ of the space group $G$ is endowed with a natural basis in the space
\begin{equation}
\label{eq:def_O}
    \mathcal{O}=\left(\bigotimes_{i=1}^{n}V_i\right)\otimes\mathbb{Z}^d,
\end{equation}
of spinful electronic orbitals throughout the crystal. 
Here $V_1$ is a vector space spanned by orbitals transforming in the representation $\rho$ of the site-symmetry group $G_\mathbf{q}=G_{\mathbf{q}_1}$; $n$ is the multiplicity of the Wyckoff position, and each of the $V_{i\neq 1}$ is the image of $V_1$ under the Wyckoff orbit. 
Finally $\mathbb{Z}^d$ represents the d-dimensional group of lattice translations. 
If $\rho_\mathbf{k}$ is an EBR, then $\rho$ is an irreducible representation (irrep) of the site symmetry group. 
The spin content of the band representation is contained within the vector spaces $V_i$; as a representation of the site symmetry group $G_{\mathbf{q}_i}$, it is a subspace
\begin{equation}
    V_i\subset L\otimes\mathbb{CP}^1,
\end{equation}
where $L$ are orbitals transforming in a spinless representation of $SO(3)$ (i.e., $s$, $p$, or $d$ orbitals), and $\mathbb{CP}^1$ is the Bloch sphere of spin.

The band representation $\rho_\mathbf{k}$ encodes the information about the symmetry properties of bands throughout the entire BZ. 
We can subduce the band representation onto the little group $G_\mathbf{k_*}$ of a high symmetry point $\mathbf{k}_*$ to find the irreps of the little group---and hence the degeneracies---that occur in the band representation, i.e.
\begin{equation}
\rho_k\downarrow G_{\mathbf{k}_*} =\bigoplus_i \eta_i,
\end{equation}
where the $\eta_i$ are irreps of the little group $G_\mathbf{k_*}$.  
In principle, the $\eta_i$ can be determined using Schur orthogonality relations for the little group and site symmetry group representations, without ever referring to the basis $\mathcal{O}$ of the band representation\cite{Cracknell,elcoro2017double}. 
However, if we keep track of the basis $\mathcal{O}$, then the band representation gives us a map
\begin{equation}
s^{\rho_\mathbf{k}}_{i\mathbf{k_*}}:W_{i\mathbf{k}_*}\rightarrow \mathcal{O},
\end{equation}
from the space $W_{i\mathbf{k_*}}$ carrying the irrep $\eta_i$ of the little group into the basis $\mathcal{O}$ of the band representation. 
Concretely, the ``pullback map'' $s^{\rho_\mathbf{k}}_{i\mathbf{k_*}}$ is a $(n\times\mathrm{dim}(V_1))\times\mathrm{dim}(W_i)$ matrix that reexpresses the states spanning the degeneracy $W_i$ in terms of the original spin and orbital basis states.

Going further, we can examine how bands disperse away from the degeneracies at $\mathbf{k}_*$. 
For small deviations, we can use symmetry-constrained $\mathbf{k}\cdot\mathbf{p}$ perturbation theory to write an effective $\dim(W_i)\times\dim(W_i)$ Bloch Hamiltonian describing the dispersion of the degeneracy $\eta_i$ away from $\mathbf{k}_*$. 
If $|m\mathbf{k}\rangle\in W_i$ is an eigenstate of the $\mathbf{k}\cdot\mathbf{p}$ Hamiltonian of the degeneracy $\eta_i$ at $\mathbf{k}$ near $\mathbf{k_*}$, then we can define pullback state
\begin{equation}
\label{eq:def_pullback}
    |m\mathbf{k}(vn\sigma)\rangle=(s_{i\mathbf{k}_*}^{\rho_\mathbf{k}})|m\mathbf{k}\rangle\in \mathcal{O},
\end{equation}
which expresses the eigenstates of the $\mathbf{k}\cdot\mathbf{p}$ Hamiltonian near $\mathbf{k}_*$ in the basis of the band representation, where $v$ indexes states in $V_i$, $n$ indexes the $V_i$, and $\sigma$ is the spin. In the remainder of this work, we will often leave the map $s$ implicit in cases where no confusion will arise. 
The pullback state encodes how the spin and orbital degrees of freedom for the Bloch eigenstates evolve as a function of $\mathbf{k}$.

Now assume we have a measurement scheme like spin-ARPES, which can measure the spin of an electronic state, but is insensitive to orbital degrees of freedom. 
Then the spin detected in a state $|m\mathbf{k}\rangle$ near a degeneracy $\eta_i$ is described by the reduced density matrix
\begin{align}
    M_S(\mathbf{k})&=\mathrm{Tr}_{\mathrm{orbitals}}\left[(s_{i\mathbf{k}_*}^{\rho_\mathbf{k}})|m\mathbf{k}\rangle\langle m\mathbf{k}|(s_{i\mathbf{k}_*}^{\rho_\mathbf{k}})^{T}\right]\nonumber\\
    &=\frac{1}{2}(\sigma_0+\mathbf{n}(\mathbf{k})\cdot\sigma).
\end{align}
A short computation shows that ${\bf n}({\bf k})=(\langle \sigma_x\rangle, \langle \sigma_y \rangle, \langle \sigma_z \rangle)$ gives ($2/\hbar$ times) the average spin vector that would be measured in an experiment.
The magnitude of the vector $\mathbf{n}(\mathbf{k})$ quantifies the ``quality" of the spin momentum locking, where $|\mathbf{n}(\mathbf{k})|=1$ corresponds to a pure spin state, and $|\mathbf{n}(\mathbf{k})|=0$ is a maximally mixed (unlocked) state. 
The unit vector $\mathbf{\hat{n}}(\mathbf{k})$ gives the spin-locking direction. 
We thus have a map between band representations, and spin vectors $\mathbf{n}(\mathbf{k})$ near degenerate points. 

Note that the pullback maps $s_{i\mathbf{k}_*}^{\rho_\mathbf{k}}$---and hence the sRDM vectors $\mathbf{n}(\mathbf{k})$ depend on the specific basis used for the band representation, rather than just its representation-theoretic content. 
This suggests that $\mathbf{n}(\mathbf{k})$ can change continuously as a function of perturbations such as the energy splitting between bands carrying the same irreps. 
Nevertheless, as long as $|\mathbf{n}(\mathbf{k})|\neq 0$, we can define an integer-valued winding number
\begin{equation}
\label{eq:chern_number}
    \nu=\frac{1}{8\pi}\int{dS^\mu\epsilon_{\mu\nu\lambda}\mathbf{n}\cdot\frac{\partial \mathbf{n}}{\partial k_\nu}\times\frac{\partial \mathbf{n}}{\partial k_\lambda}},
\end{equation}
where the integral is taken over a constant energy surface with normal vector $d\mathbf{S}$, $\epsilon_\mu\nu\lambda$ is the antisymmetric Levi-Civita symbol, $\mu,\nu,\lambda$ index the three momentum directions, and repreated indices are summed. 
As an integer invariant, the winding number $\nu$ characterizes the topology of the spin texture $\mathbf{n}(\mathbf{k})$ and is robust to small perturbations. 

In the remainder of this work, we will explore how to compute the pullback maps $s_{i\mathbf{k}_*}^{\rho_\mathbf{k}}$ and sRDM vectors $\mathbf{n}(\mathbf{k})$ from topological quantum chemistry. 
We will focus on two representative examples: the multiplicity-one Wyckoff positions in space group $P432$ (207), and the multiplicity-four Wyckoff positions in space group $P2_13$ (198). 
By focusing on exactly-solvable limits of the $\mathbf{k}\cdot\mathbf{p}$ Hamiltonian near the degeneracies that arise, we will be able to compute the winding number for the resulting spin textures, which we expect to be robust even away from the exactly-solvable limit.

\section{Band representations from multiplicity-one Wyckoff positions}
\label{sec:1apos}

In this section, we shall use the approach outlined above to calculate the sRDMs for space group $P432$ (207). 
More specifically, we shall consider the  band representations $\rho_{\bf k}=(\rho\uparrow G)_\mathbf{q}$ at multiplicity-one Wyckoff positions.  
In this case, the band representation $\mathcal{O}=V\otimes\mathbb{Z}^d $ is spanned by the spinful basis functions $v\in V$ of the irrep $\rho$ of the site-symmetry group, repeated in each unit cell.
Despite its simple nature, the procedure outlined here will be applicable to more complex cases, and the results will serve as building blocks for the sRDM of systems with high-multiplicity Wyckoff positions, as we will see in Sec.~\ref{sec:Band Representations from high-multiplicity Wyckoff Positions}.

\subsection{Weyl fermion with s-like orbitals}

Space group $P432$ (207) is a symmorphic space group with a cubic Bravais lattice and chiral octahedral point group $432$. 
The point group (viewed as a subgroup of the space group) has two independent generators, which we take to be $\left\{C_{31}^+|000\right\}$ and $\left\{C_{2a}|000\right\}$, which are respectively the three-fold rotation around the $(111)$ axis, and the twofold rotation with respect to the $(110)$ axis. 
The site symmetry group of multiplicity-one Wyckoff positions is isomorphic to the point group. 
In Appendix~\ref{sec:Some Useful Character Tables} we summarize the irreps and basis functions for the point group $432$. 

Let us first consider spinful $s$-orbitals $|j,m\rangle=|1/2,\pm1/2\rangle$ with total angular momentum $j=1/2$ at a multiplicity-one Wyckoff position (either $1a$ or $1b$). 
In the $|j,m\rangle$ basis, the generators of the point group are represented as
\begin{eqnarray}\begin{aligned}
\label{eq:rep_j}
\rho^j(C_{31}^+) &=e^{-i\frac{2\pi}{3\sqrt{3}}(J^{j}_x+J^{j}_y+J^{j}_z)},\\
\rho^j(C_{2a}) &=e^{-i\frac{\pi}{\sqrt{2}}(J^{j}_x+J^{j}_y)},
\end{aligned}\end{eqnarray}
where ${\bf J}^j$ are the vector of rotation generators for states with total angular momentum $j$. 
By comparison with Table.~\ref{tab:432chars}, it is straightforward to verify that the $j=1/2$ states carry a two-dimensional irrep of the site symmetry group labelled by $\bar{E}_1$. 

Since we have only one site per unit cell, the induction to the band representation $\rho^j_\mathbf{k}=(\rho^j\uparrow G)$ is rather trivial; at each time-reversal invariant momentum (TRIM) $\mathbf{k}_*$, we find that $\rho^j_\mathbf{k}\downarrow G_{\mathbf{k}_*}$ yields a two-dimensional representation.  
For each twofold degeneracy, we can examine its $\mathbf{k}\cdot\mathbf{p}$ Hamiltonian. 
Each symmetry generator $g$ of the group will place a constraint on the possible form of the Hamiltonian as 
\begin{eqnarray}\begin{aligned}
\label{eq:constriants_eqn}
\rho_{\bf k_*}^{1/2}(g)H(\mathbf{k})\rho_{\bf k_*}^{1/2}(g)^{-1} &=H({g{\bf k}}).\\
\end{aligned}\end{eqnarray} 
With these constraints, for $j=1/2$, we can solve the most general Hamiltonian for the $s$-orbital as
\begin{eqnarray}\begin{aligned}
\label{eq:H_two_fold}
    H^{(2)}({\bf\delta k})=v(k-k_*)_iJ^{1/2}_i\equiv v_f{\bf \delta k}\cdot {\bf J}^{1/2},
\end{aligned}\end{eqnarray}
for any TRIM ${\bf k}_*$, where the Fermi velocity $v_f$ cannot be determined by symmetry alone. 
Since we are starting from a multiplicity-one Wyckoff position, the form of the invariant Hamiltonian depends on neither the specific TRIM nor the specific choice of multiplicity-one position: both choices will multiply the representation matrices $\rho^j$ by overall phases, which do not affect the commutation relations with the $\mathbf{k}\cdot\mathbf{p}$ Hamiltonian.
The eigenstate of $H^{(2)}({\bf\delta k})$ can be written as
\begin{eqnarray}\begin{aligned}
\label{eq:Eigenstates_two_fold}
|\delta\mathbf{k}m\rangle=e^{-i\phi J_z^{1/2}}e^{-i\theta J_y^{1/2}}\ket*{\frac{1}{2},m},
\end{aligned}\end{eqnarray}
where $\ket{1/2,m}$ are the eigenstates of $J^{1/2}_z$ and we have defined the angles (we will use this convention throughout)
\begin{align}
    \cos\theta=\frac{\delta k_z}{|\delta\mathbf{k}|} , \quad
    \tan\phi=\frac{\delta k_y}{\delta k_x}.
\end{align}
Using this, we can look at the sRDM for a constant energy surface, to determine the spin-momentum locking. 
However, since we have trivial orbital degrees of freedom, the reduced density matrix here is just the ordinary density matrix. 
We find that $M_S^\pm(\mathbf{\delta k})=|{\bf \delta k},m\rangle\langle{\bf \delta k},m|=\frac{1}{2}(\sigma_0\pm\hat{\mathbf{{\delta k}}}\cdot\sigma)$, or 
\begin{equation}
\label{eq:spinRDM_s}
   \mathbf{n}_{1/2}^{\pm1/2}(\delta\mathbf{k})=\pm\hat{\delta\mathbf{k}}.
\end{equation}
There is thus perfect spin-momentum locking in this twofold degeneracy, and the winding number
\begin{equation}
\nu^{m=\pm1/2}_{j=1/2} =\pm 1
\end{equation}
by Eq.~\eqref{eq:chern_number}.

\subsection{Fourfold degeneracy with higher Orbitals}
\label{sec:Fourfold degeneracy with higher Orbitals}

We can perform a similar analysis for  the fourfold degeneracies at the $\Gamma$ and $R=(\pi,\pi,\pi)$ points in space group $P432$ (207), which are subduced in band representations with multiplicity-one Wyckoff positions.
Let us first consider the band representation carried by the $p$-orbitals located at either the $1a$ or $1b$ position. 
Under the action of the site symmetry group, the $p$-orbitals transform in the (reducible) representation given by
\begin{align}
 \rho^p(C_{31}^+) &=e^{-i\frac{2\pi}{3\sqrt{3}}(J^{1}_x+J^{1}_y+J^{1}_z)}\otimes e^{-i\frac{2\pi}{3\sqrt{3}}(J^{1/2}_x+J^{1/2}_y+J^{1/2}_z)},\nonumber\\
\rho^p(C_{2a}) &=e^{-i\frac{\pi}{\sqrt{2}}(J^{1}_x+J^{1}_y)}\otimes e^{-i\frac{\pi}{\sqrt{2}}(J^{1/2}_x+J^{1/2}_y)},
\end{align}
 where the first and second factor are for the spinless $p$-orbital and the spin part respectively. 
 In this basis, the states are spin-orbit decoupled and we shall denote them as $|j_1,m_1;j_2,m_2\rangle$ where $j_1=1$ for the $p$-orbital and $j_2=1/2$ for spin. 
 Consulting Table~\ref{tab:432chars}, we can identify this as the $T_1\otimes \bar{E}_1$ representation of the point group. 
 By taking traces of the representation matrices, we find that this representation splits as
 \begin{equation}
 T_1\otimes\bar{E}_1 = \bar{E}_1 \oplus \bar{F},
 \end{equation}
where $\bar{E}_1$ is the two-dimensional $j=1/2$ representation we encountered for $s$-orbitals, and $\bar{F}$ is the four dimensional $j=3/2$ representation. 
To find the base for these representations in terms of our spinful $p$-orbitals, we take inspiration from the theory of addition of angular momentum. 
Since the states in the degenerate subspaces have total angular momentum $j=1/2,3/2$, the basis states can be labeled as $|j,m\rangle$ respectively (where, unless otherwise specified, the spin quantization axis is taken to be $\mathbf{\hat{z}}$).  These states are inherently strongly spin-orbit coupled. We can decompose them into linear combinations of  spin-orbit decoupled states as
\begin{align}
\label{eq:CG_coe}
|j,m\rangle &= \sum_{m_1=-j_1}^{j_1}\sum_{m_2=-j_2}^{j_2}|j_1,m_1;j_2,m_2\rangle\langle j_1,m_1;j_2,m_2|j,m\rangle,
\end{align}
where $\langle j_1,m_1;j_2,m_2|j,m\rangle$ are Clebsch-Gordan coefficients. 
The Clebsch-Gordan coefficients serves as a unitary transformation to block-diagonalize the site symmetry group representations as
\begin{align}
  \rho^p(C_{31}^+) &=e^{-i\frac{2\pi}{3\sqrt{3}}(J^{\frac{1}{2}}_x+J^{\frac{1}{2}}_y+J^{\frac{1}{2}}_z)}\oplus e^{-i\frac{2\pi}{3\sqrt{3}}(J^{\frac{3}{2}}_x+J^{\frac{3}{2}}_y+J^{\frac{3}{2}}_z)},\nonumber\\
\rho^p &=e^{-i\frac{\pi}{\sqrt{2}}(J^{\frac{1}{2}}_x+J^{\frac{1}{2}}_y)}\oplus e^{-i\frac{\pi}{\sqrt{2}}(J^{\frac{3}{2}}_x+J^{\frac{3}{2}}_y)},
\end{align}
where the summands correspond to the irreps carried by $P_{1/2}$ and $P_{3/2}$ orbitals respectively. 

For each irrep, we can induce the band representation and subduce to either the $\Gamma$ or $R$ point. 
As for $s$-orbitals, the induction procedure is rather trivial, and we arrive at the constraint equation
\begin{equation}\label{eq:constraint_eqn_more_general}
\rho^{j}(g)H(\mathbf{k})\rho^{j}(g)^{-1} =H({g{\bf k}}),
\end{equation}
for $j=1/2,3/2$. 
For the twofold degeneracy, the most general Hamiltonian is again given by Eq.~\eqref{eq:H_two_fold} with the eigenstates given in Eq.~\eqref{eq:Eigenstates_two_fold}. 
However, 
unlike the case of $s$-orbital, where the spin-orbit coupled state is essentially the same as the decoupled state, for the $P_{1/2}$ orbital, the Clebsch-Gordan coefficients serve as a nontrivial pull-back map from the spin-orbit coupled state to the decoupled state. 
Thus,
\begin{align}
\label{eq:207_eigvec_P1/2}
    |\delta\mathbf{k}m\rangle
    &=\sum_{m_1=-j_1}^{j_1}\sum_{m_2=-j_2}^{j_2}|j_1,m_1;j_2,m_2\rangle\times\nonumber\\
    &\quad \braket*{j_1,m_1;j_2,m_2\bigg\vert e^{-i\phi J_z^{1/2}}e^{-i\theta J_y^{1/2}}}{\frac{1}{2},m},
\end{align}
for $j_1=1,j_2=1/2$. 
Eq.~\eqref{eq:207_eigvec_P1/2} is a special case of Eq.~\eqref{eq:def_pullback} as the pull-back map; the second line of Eq.~\eqref{eq:207_eigvec_P1/2}, is nothing but the Clebsch-Gordan coefficients in Eq.~\eqref{eq:CG_coe}. 
We proceed to form the density matrix $|\delta\mathbf{k}m\rangle\langle\delta\mathbf{k}m|$ and trace out the orbital degree of freedom to calculate the sRDM as
\begin{equation}
\label{eq:p12locking}
 {\bf n}_{1/2}^{\pm1/2}(\delta\mathbf{k}) = \mp\frac{\hat{\bf \delta k}}{3}.
\end{equation}
Unlike for $s$-orbitals, we see that the sRDM for the Kramers Weyl fermion with $P_{1/2}$ orbitals exhibits imperfect spin-momentum locking, with $|\mathbf{n}^{\pm1/2}_{1/2}(\mathbf{k})|=1/3 < 1$. 
From this example, it is clear that Kramers Weyls may exhibit imperfect spin-momentum locking even in isolated two-band models, provided the bands are strongly spin-orbit coupled\cite{chang2018topological}. 
Further, we note that spin-momentum locking, which serves as an indicator of entanglement between different elementary band representations, is determined not only by the band representation, but also by the form of the basis functions. 
Nevertheless, upon evaluating the winding number, we see that
\begin{equation}
\nu^{m=\pm1/2}_{j=1/2} =\mp 1
\end{equation}
just as for $s$ orbitals.

We next calculate the sRDM for the fourfold degeneracy subduced by $P_{3/2}$ orbitals at $\Gamma$ and $R$. 
Upon solving the constraint equation in Eq.~\eqref{eq:constraint_eqn_more_general}, we  find that the most general Hamiltonian for the fourfold degeneracy reads\cite{Bradlyn2016,flicker2018chiral} 
\begin{widetext}
\begin{equation}
\label{eq:H_207_P_3/2}
    H^{(4)}({\bf \delta k})=
    \hbar v\left(\begin{array}{cccc}
    ak_z & \frac{\sqrt{3}(a+b)}{4}k_- & 0 & \frac{a-3b}{4}k_+  \\
    \frac{\sqrt{3}(a+b)}{4}k_+ & bk_z & \frac{3a-b}{4}k_- & 0 \\
    0 & \frac{3a-b}{4}k_+ & -bk_z & \frac{\sqrt{3}(a+b)}{4}k_- \\
    \frac{a-3b}{4}k_- & 0 & \frac{\sqrt{3}(a+b)}{4}k_+ & -ak_z
    \end{array}\right),
\end{equation}
\end{widetext}
where 
$a=\cos\chi$, $b=\sin\chi$ and $k_\pm = \delta k_x\pm i\delta k_y$. 
While this Hamiltonian is hard to diagonalize for generic values of $v$ and $\chi$, there exist several exactly solvable points. 
Focusing on the case where $\chi=\arctan(1/3)$, we have $H\propto \delta\mathbf{k}\cdot \mathbf{J}^{3/2}$, with eigenstates
\begin{eqnarray}\begin{aligned}
\label{eq:207_eigvec_P3/2}
|\delta\mathbf{k}m\rangle&=e^{-i\phi J_z^{3/2}}e^{-i\theta J_y^{3/2}}|\frac{3}{2},m\rangle\\
&=\sum_{m_1=-j_1}^{j_1}\sum_{m_2=-j_2}^{j_2}|j_1,m_1;j_2,m_2\rangle\\
    &\quad \times\braket*{j_1,m_1;j_2,m_2\bigg\vert e^{-i\phi J_z^{3/2}}e^{-i\theta J_y^{3/2}}}{\frac{3}{2},m},
\end{aligned}\end{eqnarray}
where in the second line we have decomposed the states to the spin-orbit decoupled states with the Clebsch-Gordan coefficients. 
Upon tracing out the orbital degrees of freedom for the density matrix $|\delta\mathbf{k}m\rangle\langle\delta\mathbf{k}m|$, we find that 
\begin{equation}
\label{eq:p32locking1}
        {\bf n}_{3/2}^{\pm3/2}(\delta\mathbf{k})=\pm\hat{\bf \delta k} ,
\end{equation}
indicating that the states with $m=\pm 3/2$ are perfectly spin-momentum locked. 
On the other hand, 
\begin{equation}
\label{eq:p32locking2}
    {\bf n}_{3/2}^{\pm1/2}(\delta\mathbf{k}) = \pm\frac{\hat{\bf \delta k}}{3},    
\end{equation}
which indicates imperfect spin-momentum locking of these bands.
The corresponding winding numbers for these sRDMs are found to be 
\begin{align}
\nu_{j=3/2}^{m}&=\text{sgn}(m)
\end{align}
which is $\pm1$ depending on the sign of the magnetic quantum numbers. 
Eqs.~\eqref{eq:p32locking1} and \eqref{eq:p32locking2} are obtained for $H^{(4)}(\delta{\bf k})$ specifically tuned to a exactly solvable point, and we expect the sRDMs will change if we move away from these limits. 
Nevertheless, the winding numbers, which are quantized, will not change as long as both $|\mathbf{n}_j^m(\mathbf{k})|>0$ and the bands remain nondegenerate over the Fermi surface. 
We note also that when $\chi=-\pi/4$, then after a reordering of the basis we can put Eq.~\eqref{eq:H_207_P_3/2} into the form $H\propto \delta\mathbf{k}\cdot(\mathbf{J}^{3/2})^*$\cite{Bradlyn2016}. 
The sRDM and winding number in this limit can be obtained from Eqs.~\eqref{eq:207_eigvec_P3/2}--\eqref{eq:p32locking2} by the substitution $\phi\rightarrow -\phi$ (or equivalently, $\delta k_y \rightarrow -\delta k_y$). 
Away from the $\mathbf{\delta k}\cdot\mathbf{J}$ limit, our formalism can still be applied using the eigenstates of $H^{(4)}$ obtained from numerical diagonalization.

In summary, we have 
outlined a general procedure to calculate the sRDM, and 
exhibited the results for $s$ and $p$-orbitals for the space group 207. 
One key feature demonstrated here is that for a single spinless orbital, we could end up with either an elementary band representation (like with $s$ orbitals), or a composite band representation (like with $p$ orbitals). 
We saw that whether the band representation was elementary or composite affected the quality of the spin-momentum locking, measured by $|\mathbf{n}(\mathbf{k})|$. 
Further, 
in a band structure with both $s$ and $p$ orbitals at the $1a$ position, a generic twofold degeneracy will be spanned by some linear combination of pseudospin-1/2 states from each band representation. 
The most general reduced density matrix will then be a convex linear combination of those from each band representation in the spectrum, as we will explore in Sec.~\ref{sec:composite}.

\section{Composite band representations}
\label{sec:Composite band representations}

In this section, we shall explore in more detail the sRDM in composite band representations.
As we saw in Sec.~\ref{sec:1apos}, composite band representations can affect the quality of the spin-momentum locking in an EBR. 
There are two ways this can occur. 
The first, which we shall call type-I entanglement, is when a single type of spinful orbital induces a composite band representation due to on-site spin-orbit coupling. 
The second type of entanglement occurs when  band representations from orthogonal sets of spinless orbitals subduce the same degeneracy at  ${\bf k}_*$. 
We shall explain in details the two types of entanglement. 
For simplicity, we will compute the sRDMs below only for the simplest $\delta\mathbf{k}\cdot\mathbf{J}^{j}$ form of the $\mathbf{k}\cdot\mathbf{p}$ Hamiltonians. 
Nevertheless, the sRDM for less fine-tuned Hamiltonians can be computed numerically using our same method.

\subsection{Type-I entanglement: degeneracies with non-minimal orbital content}
\label{sec:Type-I entanglement: degeneracies with non-minimal orbital content}

In Sec.~\ref{sec:Fourfold degeneracy with higher Orbitals}, we calculated the sRDM from the fourfold degeneracy of $P_{3/2}$ orbitals, and the twofold degeneracy of $P_{1/2}$ orbitals in space group $P432$ (207). 
The fourfold degeneracy is inherently strongly spin-orbit coupled, however to define the basis states we started from a spin-orbit decoupled basis. 
In order to do so, we relied on strong on-site spin orbit coupling to separate $P_{3/2}$ orbitals from $P_{1/2}$ orbitals at each site. 
In other words, to obtain the fourfold degeneracy from electrons with separate spin and orbital degrees of freedom, our Hilbert space must be at least six dimensional. 
In fact, the band representation $(\bar{F}\uparrow G)_{1a}$ with its fourfold degeneracy is naturally accompanied by the band representation $(\bar{E}_1\uparrow G)_{1a}$ carried by $P_{1/2}$ orbitals. 
We saw that this entanglement resulted in $|\mathbf{n}^{\pm1/2}_j|=1/3<1$ for both the twofold and fourfold degeneracies subduced by the $p$ orbitals.

As we saw above, on-site spin-orbit entanglement is expected to arise whenever a site-symmetry group representation is not an irreducible tensor product of spin and orbital degrees of freedom. 
In other words, on-site SOC can affect the purity of spin states - an $S_{1/2}$ state is a pure spin state, while a $P_{1/2}$ state is not. 
As we saw in the case of space group $P432$ (207), this means that, for the purpose of spin-momentum locking, we should consider \emph{single-valued} EBRs, which are doubled due to the additional spin degree of freedom. 
The additional entanglement due to spin-orbit coupling is then explained because a spin-doubled copy of a single-valued EBR needs not be a double-valued EBR (it is in general a composite band representation). 
For instance, we saw that to consider $p$ orbitals at the $1a$ position in SG 207, we have 
\begin{equation}
\label{eq:type_I_ent}
    T_1\otimes\bar{E}_1\approx\bar{E}_1\oplus\bar{F},
\end{equation}
where $T_1$ is the threefold degenerate representation of the site-symmetry group and $\bar{E}_1$ is the representation for the spin degree of freedom. 
We refer to this as ``type-I entanglement.'' We see that it cannot be removed by increasing the energy separation between the $P_{1/2}$ and $P_{3/2}$ states.

This is certainly not unique for $p$-orbitals, and in fact $d$-orbitals exhibits more interesting type-I entanglement. 
Unlike with $p$ orbitals, when we ignore spin $d$-orbitals form a \emph{reducible} representation of the point group of space group $P432$ (207) (which is the site-symmetry group of the $1a$ and $1b$ positions). 
In order to determine the appropriate basis for the band representations carried by the $d$-orbitals then, we need to consider the relative strength of on-site SOC and (spin-independent) crystal-field splitting that determines the splitting of the spinless representations.  

Let us first suppose that on-site SOC dominates. 
In this case, the ten $d$-orbitals first decouple into a $D_{3/2}$ and a $D_{5/2}$ representaton of $SU(2)$. 
The $D_{3/2}$ representation is the irreducible $\bar{F}$ representation of the site symmetry group, while the $D_{5/2}$ representation splits under the crystal field into the sum $\bar{F}\oplus\bar{E}_2$. 
We can find the bases for these representations, and hence the sRDM at the $\Gamma$ or $R$ points following the same treatment as in Sec.~\ref{sec:Fourfold degeneracy with higher Orbitals}. 
The Clebsch-Gordan coefficients in Eq.~\eqref{eq:CG_coe} block-diagonalize the  $d$-orbitals into the four-dimensional $D_{3/2}$ and six dimensional $D_{5/2}$ representations, respectively. 
The $D_{3/2}$ orbital will subduce the same little group representation that gives the same eigenstates as in Eq.~\eqref{eq:207_eigvec_P3/2}. 
The only difference is that the Clebsch-Gordan coefficients will take different values, and with these taken into account, we have
\begin{eqnarray}\begin{aligned}
   {\bf n}^{\pm3/2}_{3/2}(\delta\mathbf{k})&=\mp\frac{3}{5}{\hat{\delta\bf k}}, \quad 
    {\bf n}^{\pm1/2}_{3/2}(\delta\mathbf{k})=\mp\frac{1}{5}\hat{{\delta\bf k}},
\end{aligned}\end{eqnarray}
for the sRDM of $D_{3/2}$ orbitals. 

The situation for $D_{5/2}$ is slightly more complicated.  Upon substituting $j=5/2$ into Eq.~\eqref{eq:rep_j}, we find as mentioned above that 
the $D_{5/2}$ orbitals carry a reducible representation of the site symmetry group.  Working in the basis of angular momentum eigenstates $|5/2,\overline{m}\rangle$ for $-5/2\leq \overline{m}\leq 5/2$, the site-symmetry group representation is block-diagonal, with
\begin{align}
\rho^{D_{5/2}}(C^+_{31}) &=
e^{-i\frac{2\pi}{3\sqrt{3}}(-J^{\frac{1}{2}}_x+J^{\frac{1}{2}}_y-J^{\frac{1}{2}}_z)}\oplus  e^{-i\frac{2\pi}{3\sqrt{3}}(J^{\frac{3}{2}}_x+J^{\frac{3}{2}}_y+J^{\frac{3}{2}}_z)},\nonumber
\\
\rho^{D_{5/2}}(C_{2a})&=
e^{i\frac{2\pi}{2\sqrt{2}}(-J^{\frac{1}{2}}_x+J^{\frac{1}{2}}_y)}\oplus e^{i\frac{2\pi}{2\sqrt{2}}(J^{\frac{3}{2}}_x+J^{\frac{3}{2}}_y)}.
\end{align}
Thus we see that the $D_{5/2}$ induces a composite band representation itself, which is very different from the case of $P_{3/2}$ and $D_{3/2}$ orbitals. 

With these understood, we proceed to solve for the eigenstates of the $\mathbf{k}\cdot\mathbf{p}$ Hamiltonians subduced at $\Gamma$ and $R$. 
We have a twofold and a fourfold degeneracy which, in the $\delta\mathbf{k}\cdot\mathbf{J}$ limit have eigenstates
\begin{eqnarray}\begin{aligned}
&    |\delta\mathbf{k}m\rangle=e^{-i\phi J_z^{j}}e^{-i\theta J_y^{j}}|j,m\rangle\\
    &=\sum_{\overline{m}=-5/2}^{5/2}\sum_{m_1=-j_1}^{j_1}\sum_{m_2=-j_2}^{j_2}|j_1,m_1;j_2,m_2\rangle\times\nonumber\\
    &\quad \braket*{ j_1,m_1;j_2,m_2}{\frac{5}{2},\overline{m}}\braket*{\frac{5}{2},\overline{m}}{ e^{-i\phi J_z^{j}}e^{-i\theta J_y^{j}}\bigg\vert j,m},
\end{aligned}\end{eqnarray}
where $j=1/2,3/2$ for the twofold and fourfold degeneracy respectively.
We have explicitly written the states in the spin-orbit decoupled basis, where the pull-back map consists both the Clebsch-Gordan coefficients $ \langle j_1,m_1;j_2,m_2|5/2,\overline{m}\rangle$,  as well as the overlap integrals $\langle5/2,\overline{m}|j,m\rangle$. 
The fact that the pullback map is a composite map is a signature of the composite band representations. 
With these taken into account, for the sRDM, we find that 
\begin{eqnarray}\begin{aligned}
    {\bf n}_{1/2}^{\pm1/2}(\delta\mathbf{k}) = \pm\frac{\hat{\bf \delta k}}{3},
\end{aligned}\end{eqnarray}
for the twofold degeneracy, and
\begin{eqnarray}\begin{aligned}
\label{eq:srdm_5}
  {\bf n}^{\pm3/2}_{3/2}(\delta\mathbf{k})&=\pm\left(\frac{2}{5}\hat{\bf \delta k}-\frac{3}{5}\hat{\bf \delta k}^3\right), \\
    {\bf n}^{\pm1/2}_{3/2}(\delta\mathbf{k})&=\mp\left(\frac{16}{15}\hat{\bf \delta k}-\frac{9}{5}\hat{\bf \delta k}^3\right),
\end{aligned}\end{eqnarray}
for the fourfold degeneracy, where we have introduced the shorthand
\begin{equation}
\hat{\delta\mathbf{k}}^3 = \frac{1}{|\delta\mathbf{k}|^3}(\delta k_x^3,\delta k_y^3, \delta k_z^3).
\end{equation}
These sRDMs are labelled by ``SOC'' in Table.~\ref{tab:207}, referring to the fact that they are obtained in the limit that on-site spin-orbit coupling is large. 
Although not covariant under continuous rotations, vectors of the form $a\hat{\mathbf{\delta k}}+b\hat{\mathbf{\delta k}}^3$ transform in the vector $T_1$ representation of the chiral octahedral group (see Appendix~\ref{sec:Some Useful Character Tables}), and are hence allowed by the symmetry of space group $P432$ (207). 
We note that the winding numbers of the sRDMs in Eq.~\eqref{eq:srdm_5} are 
\begin{align}
\nu_{3/2}^{\pm3/2}&=\mp5 \\
\nu_{3/2}^{\pm1/2}&=\pm5
\end{align}
which is very different from what we found in Sec.~\ref{sec:Fourfold degeneracy with higher Orbitals} for $P_{3/2}$ orbitals. 
This shows that the winding number of the sRDM is not uniquely related to the Chern number of the pseudospin states (equal to $2m$for a ). 

Alternatively, we could consider a situation in which crystal-field splitting dominates over on-site SOC. 
In that case, the $d$ orbitals first split under the crystal field into a sum of two spinless irreps of the site-symmetry group: the three-dimensional pseduovector representation $T_2$ carried by the $(d_{xy},d_{xz},d_{yz})$ orbitals, and the two-dimensional $E$ representation carried by the $(d_{x^2-y^2},d_{z^2})$ orbitals. 
Including SOC, the $T_2$ representation decomposes as
\begin{equation}
    T_2\otimes\bar{E}_1 = \bar{E}_2\oplus \bar{F},
\end{equation}
while the two-dimensional $E$ representation yields
\begin{equation}
    E\otimes\bar{E}_1 = \bar{F}.\label{eq:dorbxtal}
\end{equation}
Although we end up with the same representations as in the strong SOC case, accounting for the crystal field splitting in this way gives us a different basis for the band representations. 
In particular, note that the four dimensional $\bar{F}$ representation from the $(d_{x^2-y^2},d_{z^2})$ orbitals is spanned by pure spin states, by virtue of Eq.~\eqref{eq:dorbxtal}. 
This is in contrast to the $D_{3/2}$ and $D_{5/2}$ orbitals in the strong-SOC limit, neither of which are pure spin states. 
We can follow the same procedure as above to compute the sRDM from the basis of these band representations; the results are summarized in the rows labelled ``Crystal field'' in Table~\ref{tab:207}. 
Crucially, we note that although the basis states for the fourfold degeneracy with $(d_{x^2-y^2},d_{z^2})$ orbitals are pure spin states, the sRDM for these states is not trivial and isotropic. 
This is because unlike the $P_{3/2}$ or $D_{3/2}$ orbitals, $(d_{x^2-y^2},d_{z^2})\otimes\{\ket{\uparrow},\ket{\downarrow}\}$ orbitals do not form a representation of $SU(2)$. 

We can work out the sRDM explicitly in this case. 
Let us choose the basis
\begin{align}
\ket{\frac{3}{2}} &= \ket{d_{x^2-y^2}\uparrow} \label{eq:ket3}\\
\ket{\frac{1}{2}} &=\ket{d_{z^2}\downarrow} \\
\ket{-\frac{1}{2}} &=\ket{d_{z^2}\uparrow} \\
\ket{-\frac{3}{2}} &=\ket{d_{x^2-y^2}\downarrow}.\label{eq:ketm3}
\end{align} 
The states are labelled by their $J_z$ eigenvalue, in a basis where 
\begin{equation}
J_z = \begin{pmatrix}
\frac{3}{2} & 0 & 0 & 0 \\
0 & \frac{1}{2} &0 & 0 \\
0 & 0 & -\frac{1}{2} & 0\\
0 & 0 & 0 & -\frac{3}{2}
\end{pmatrix}.
\end{equation}
We can find the representation matrices of the site symmetry group generators in this basis by subduction from $SU(2)$. 
We have that $C_{4z}$ is represented by
\begin{align}
\rho(C_{4z}) =\exp\left[\frac{2\pi i}{4}J_z^{3/2}\right].
\end{align}
Similarly, a direct computation shows that in this basis
\begin{equation}
\rho(C_{31}) = \exp\left[-\frac{2\pi i}{3\sqrt{3}}(J_x-J_y+J_z)\right].
\end{equation}
Finally, time-reversal symmetry in this basis is represented by
\begin{equation}
\rho(TR) = \begin{pmatrix}
0 &0 &0 &1 \\
0 & 0 & -1 & 0\\
0 & 1 & 0 &0 \\
-1 & 0 &0 & 0
\end{pmatrix}\mathcal{K}.
\end{equation}
This is \emph{almost} the usual basis for the $\bar{F}$ representation spanned by spin-$3/2$ orbitals, but with some crucial minus signs. 
It is straightforward to verify that
\begin{equation}
H=\mathbf{\delta k}\cdot(\mathbf{J}^{3/2})^* \label{eq:xfieldham}
\end{equation}
is invariant under these symmetries, in the sense of Eq.~\eqref{eq:constraint_eqn_more_general}. 
The eigenstates of this Hamiltonian are given by
\begin{equation}
\ket{m\delta\mathbf{k}} = e^{i\phi J^{3/2}_z}e^{-i\theta J_y^{3/2}}\ket{m}\label{eq:spin32rotation}
\end{equation}
where $m=\pm3/2, \pm1/2$, and $\ket{m}$ is one of the states defined in Eqs.~\eqref{eq:ket3}--\eqref{eq:ketm3}.

\begin{figure}[t]
\includegraphics[width=\columnwidth]{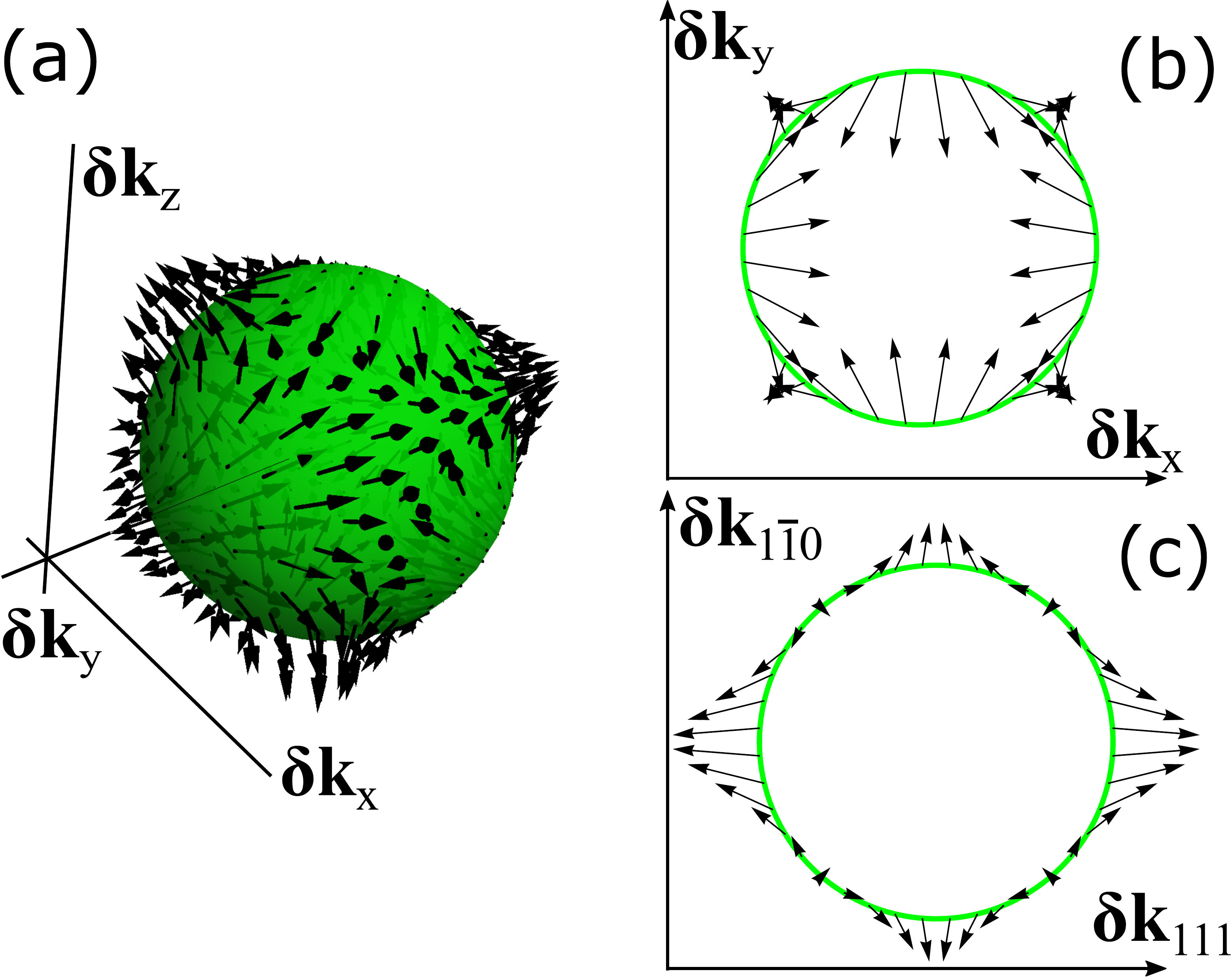}
\caption{Plot of the spin texture with winding $\nu=-5$ for crystal-field split $d$-orbitals in space group $P432$ (207). a) shows the vector $\mathbf{n}(\mathbf{k})$ on a constant-energy sphere. b) shows the projection $(n_x(\mathbf{k}),n_y(\mathbf{k}))$ of the spin texture along the projection of the constatnt energy sphere to the $k_x-k_y$ plane. c) shows the analogous projection in the $k_{111}-k_{1\bar{1}0}$ plane.}\label{fig:winding_5}
\end{figure}

From our expressions for the eigenstates, along with the fact that the states $\ket{m}$ are pure spin states, we might expect that the sRDM for this degeneracy was purely radial. 
However, it is important to note that although the continuous spin-$3/2$ rotation Eq.~\eqref{eq:spin32rotation} is a symmetry of the Hamiltonian \eqref{eq:xfieldham}, it acts nontrivially on spin degrees of freedom. 
To see this and to calculate the sRDM, we note that in the basis of $\{\ket{m}\}$ the spin operators take the form
\begin{align}
s_x &= \frac{1}{2}\begin{pmatrix} 0 & 0 & 0 & 1 \\
0 & 0 & 1 & 0 \\
0 & 1 & 0 & 0 \\
1 & 0 &0 & 0
\end{pmatrix}, \nonumber\\
s_y  &= \frac{1}{2}\begin{pmatrix} 0 & 0 & 0 & -1 \\
0 & 0 & 1 & 0 \\
0 & -1 & 0 & 0 \\
1 & 0 &0 & 0
\end{pmatrix}, \\
s_z &=\frac{1}{2}\begin{pmatrix} 1 & 0 & 0 & 0 \\
0 & -1 & 0 & 0 \\
0 & 0 & 1 & 0 \\
0 & 0 &0 & -1
\end{pmatrix}\nonumber.
\end{align}
The vector $\vec{s}$ of spin matrices do \emph{not} transform as a vector under rotations generated by the $\mathbf{J}^{3/2}$. 
Computing the sRDM, we have
\begin{align}
\mathbf{n}_m &= \bra{m\mathbf{k}}\vec{s}\ket{m\mathbf{k}}\nonumber \\
& \bra{m}e^{i\theta J_y^{3/2}}e^{-i\phi J_z^{3/2}}\vec{s}e^{i\phi J^{3/2}_z}e^{-i\theta J_y^{3/2}}\ket{m} \\
&=\begin{cases}
\pm (\hat\delta{\mathbf{k}})^3,\;\; &m=\pm 3/2 \\
\pm \left[2\hat\delta{\mathbf{k}}-3(\hat\delta{\mathbf{k}})^3\right],\;\; &m=\pm 1/2\nonumber
\end{cases}
\end{align}
as stated in Table~\ref{tab:207}. 
We can evaluate the winding number of the spin over the Fermi surface for each band to find
\begin{align}
\nu_m &= \frac{1}{4\pi}\int d\theta d\phi \hat{\mathbf{n}}_m\cdot\left[\frac{\partial\hat{\mathbf{n}}_m}{\partial \theta}\times\frac{\partial\hat{\mathbf{n}}_m}{\partial \phi}\right] \nonumber\\
& =\begin{cases} 
\pm 1,\;\; &m=\pm 3/2 \\
\mp 5,\;\; &m=\pm 1/2
\end{cases}.
\end{align}
In Fig.~\ref{fig:winding_5} we show a plot of the spin texture with winding $-5$.

\begin{table*}[htp]
\begin{center}
\begin{tabular}{|c|c|c|c|c|c|c|c|c|}
\hline
\multicolumn{2}{|c|}{} & $s$ & $p$ & \multicolumn{5}{|c|}{$d$} \\
\hline
\multicolumn{2}{|c|}{}  &  &  &&\multirow{2}{*}[0cm]{$D_{3/2}$}&\multirow{2}{*}[0cm]{fourfold}&$m=\pm\frac{1}{2}$& $\mp\frac{1}{5}\hat{\bf \delta k}~[\mp1]$ \\
\cline{8-9}
\multicolumn{2}{|c|}{}  &  &  &&&&$m=\pm\frac{3}{2}$& $\mp\frac{3}{5}\hat{\bf \delta k}~[\mp1]$ \\
\cline{6-9}
\multicolumn{2}{|c|}{twofold}  & $\pm\hat{\bf \delta k}~[\pm1]$ & $\mp\frac{1}{3}\hat{\bf \delta k}~[\mp1]$ &\parbox[t]{2mm}{\multirow{3}{*}[0.3cm]{\rotatebox[origin=c]{90}{SOC}}}&&\multicolumn{2}{|c|}{twofold}& $\pm\frac{1}{3}\hat{\bf \delta k}~[\pm1]^*$ \\
\cline{7-9}
\multicolumn{2}{|c|}{}  &  &  &&$D_{5/2}$&\multirow{2}{*}[0cm]{fourfold} &$m=\pm\frac{1}{2}$& $\mp(\frac{16}{15}\hat{\bf \delta k}-\frac{9}{5}\hat{\bf \delta k}^3)~[\pm5]$ \\
\cline{8-9}
\multicolumn{2}{|c|}{}  &  &  &&&&$m=\pm\frac{3}{2}$& $\pm(\frac{2}{5}\hat{\bf \delta k}-\frac{3}{5}\hat{\bf \delta k}^3)~[\mp5]$ \\
\hline
  &&&  && &\multicolumn{2}{|c|}{twofold}& $\pm\frac{1}{3}\hat{\bf \delta k}~[\pm1]^*$ \\
  \cline{7-9}
  &$m=\pm1/2$ &&$\pm\frac{1}{3}\hat{\bf \delta k}~ [\pm1]$  &&$d_{xy}, d_{yz}, d_{zx}$&\multirow{2}{*}[0cm]{fourfold}&$m=\pm\frac{1}{2}$& $\pm(\hat{\bf \delta k}-2\hat{\bf \delta k}^3)~[\times]$ \\
\cline{8-9}  
fourfold & & $\times$ &  & \parbox[t]{2mm}{\multirow{3}{*}[0.7cm]{\rotatebox[origin=c]{90}{Crystal field}}} & & & $m=\pm\frac{3}{2}$& $\mp(\hat{\bf \delta k}-\frac{2}{3}\hat{\bf \delta k}^3)~[\mp1]$\\
\cline{2-2}\cline{4-4}\cline{6-9}
&\multirow{2}{*}[0cm]{$m=\pm3/2$} &&\multirow{2}{*}[0cm]{$\pm\hat{\bf \delta k}~ [\pm1]$}  &&\multirow{2}{*}[0cm]{$d_{x^2-y^2}, d_{z^2}$}&\multirow{2}{*}[0cm]{fourfold}&$m=\pm\frac{1}{2}$& $\pm(2\hat{\bf \delta k}-3\hat{\bf \delta k}^3)~[\mp5]^*$ \\
\cline{8-9}
& &&  &&&&$m=\pm\frac{3}{2}$& $\pm\hat{\bf \delta k}^3~[\pm1]^*$ \\
\hline
\end{tabular}
\end{center}
\caption{sRDM of $s$, $p$ and $d$ orbitals for space group 207. $\times$ indicates that the band representation of the spinful orbital will not subduce an irrep with that dimension at ${\bf k}_*$. 
The square brakets indicate the corresponding winding numbers of the degeneracies, with $[\times]$ indicates the winding number is ill-defined for the sRDM because $|\mathbf{n}(\mathbf{k})|$vanishes at certain points on the Fermi surface. An asterisk is used to denote cases where the invariant Hamiltonian has the form $\delta\mathbf{k}\cdot\mathbf{J}^*$.
}
\label{tab:207}
\end{table*}%

We see then that a general feature of type-I entanglement is the dependence of the sRDM on the interplay between on-site SOC and crystal field splitting. 
For type-I entangled orbitals the precise basis carried by each EBR can range from maximally spin-decoupled when crystal field splitting dominates over on-site SOC, to maximally coupled when on-site SOC is sufficiently strong that the basic building blocks of band representations are total angular momentum eigenstates. 
Most realistic materials will be somewhere in between these two extreme cases. 
However, we expect that for most crystalline solids made of light- or intermediate-weight elements, crystal field splitting will dominate over on-site SOC. 
Finally, we note that type-I entanglement can also occur when a single EBR subduces two isomorphic representations at the same high symmetry point. 
In this case, group theory alone does not uniquely specify the pullback map, per Schur's lemma. 
While this situation lies outside the scope of the present work, it may open the door to even more exotic spin textures than those presented here.

\subsection{Type-II entanglement}\label{sec:composite}

In contrast, ``type-II entanglement" occurs when we add additional independent band representations which subduce the same little group irreps at $\mathbf{k}_*$, but are carried by eigenfunctions with orthogonal orbital wavefunctions. 
In this case, the basis functions for the irreps are not restricted by symmetry, but can be arbitraty linear combinations of the basis functions from each copy, due to Schur's lemma. 
In accordance with perturbation theory, the energy splitting between irreps, along with band topology, determines the relative weight of each set of basis functions. 

Let us take space group $P432$ (207) with $s$ and $p$ orbitals as an example. 
As we have seen in Sec.~\ref{sec:1apos}, the band representations of $s$ and $P_{1/2}$ orbitals, denoted as $\rho_{\bf k}^{1,2}$ respectively, both subduce a two dimensional irrep $\eta$ at the TRIM points. 
If $\{|\alpha^1_i\rangle\}\equiv\{|\alpha^1_1\rangle,|\alpha^1_2\rangle\}$ is a basis for $\eta^1$ formed by states in $\rho_\mathbf{k}^1$, and if $\{|\alpha^2_i\rangle\}$ is a basis for $\eta^2$ formed by states in $\rho_\mathbf{k}^2$, then the two degeneracies $\eta^1$ and $\eta^2$ will be spanned by the unitary linear combinations
\begin{align}
\label{eq:unitconstraint}
    \{|\beta^1_i\rangle\}&=a\{|\alpha^1_i\rangle\}+b\{|\alpha^2_i\rangle\},\nonumber\\
    \{|\beta^2_i\rangle\}&=a^*\{|\alpha^1_i\rangle\}-b^*\{|\alpha^2_i\rangle\},\\
    1&=|a|^2+|b|^2.\nonumber
\end{align}
Now we recall two facts: first, in the absence of any coupling between band representations, $\{|\alpha_i\rangle\}$ and $\{|\beta_i\rangle\}$ separately span the representation $\eta$. 
Second, since the band representations $\rho_\mathbf{k}^1$ and $\rho_\mathbf{k}^2$ are not type-I entangled, $\mathrm{Tr}_{\mathrm{orbs}}|\alpha^1_i\rangle\langle\alpha^2_j|=\mathrm{Tr}_{\mathrm{orbs}}|\alpha^2_i\rangle\langle\alpha^1_j|=0$ (i.e., the states are spanned by \emph{different} orbital degrees of freedom). 
Using these facts, we have for the reduced density matrices
\begin{align}
    {\bf n}_{\eta^1}({\bf \delta k})&=|a|^2{\bf n}_1({\bf \delta k})+|b|^2{\bf n}_2({\bf \delta k}), \nonumber\\
    {\bf n}_{\eta^2}({\bf \delta k})&=|b|^2{\bf n}_1({\bf \delta k})+|a|^2{\bf n}_2({\bf \delta k}),
\end{align}
where ${\bf n}_1({\bf \delta k})$ and ${\bf n}_2({\bf \delta k})$ are the reduced density matrices for the band representations $\rho_\mathbf{k}^1$ and $\rho_\mathbf{k}^2$ taken in isolation. 
Besides $s$ and $P_{1/2}$ orbitals, the contribution from the twofold degeneracy subduced by $d$-orbitals must also be included, 
as we have seen in Sec.~\ref{sec:Type-I entanglement: degeneracies with non-minimal orbital content}. 
Note also that in the ``Crystal field'' case, the two fourfold degeneracies subduced by $d$-orbitals can also type-II entangled with each other, since the basis $(d_{xy},d_{xz},d_{yz})$ and $(d_{x^2-y^2},d_{z^2})$ orbitals for each degeneracy are orthogonal to each other. 
This contrasts with the ``SOC'' case, where the fourfold degeneracis from the $D_{5/2}$ and $D_{3/2}$ orbitals were, by construction, type-I entangled---the orbital part of their basis functions are not necessarily orthogonal.
As a result, for the sRDM summarized in Table.~\ref{tab:207}, 
the sRDM of a certain degeneracy, either twofold or fourfold, is generally the linear combination of type-II entangled entries with orthogonal crystal-field split orbitals.

More generally, suppose we have a set of $N$ band representations $\rho_{\bf k}^I$, $I=1,...,N$ none of which are type-I entangled, and all of which subduce an irrep $\eta$ at ${\bf k}_*$. 
This could happen if we have multiple orbital types, or multiple occupied Wyckoff positions (see Sec.~\ref{sec:Band Representations from high-multiplicity Wyckoff Positions}). 
Let $\{|\alpha^I_i\rangle\}$ be the basis for $\eta^I$ formed by states in $\rho_{\bf k}^I$, then the states will be unitary linear combinations    
\begin{eqnarray}\begin{aligned}
\{|\beta^I_i\rangle\} = U_{IJ}\{|\alpha^J_i\rangle\},
\end{aligned}\end{eqnarray}
where $U_{IJ}$ is a unitary matrix, such that the sRDM are
\begin{eqnarray}\begin{aligned}
{\bf n}_{\eta^I}({\delta\bf k}) = \sum_{J}|U_{IJ}|^2{\bf n}_{\eta^J}({\delta\bf k}).
\end{aligned}\end{eqnarray}
 We see that the unitarity constraint, such as Eq.~(\ref{eq:unitconstraint}), ensures that the total reduced density matrix is a convex linear combination of the matrices computed from TQC. 
 In this way, the previous analysis generalizes to the addition of multiple band representations, and hence to realistic materials.

\section{Band Representations from high-multiplicity Wyckoff Positions}
\label{sec:Band Representations from high-multiplicity Wyckoff Positions}

Let us now move on to the experimentally exciting case of band representations induced from Wyckoff positions with more than one site. 
In this case, we will see that there can be non-perfect spin-momentum locking even when there is no on-site spin-orbit entanglement (i.e., when the band representation comes from $s$ orbitals). 
As we will see, this emerges due to entanglement between spin and linear combinations of orbitals at different sites in the Wyckoff position.

For concreteness, let us focus on band representations in space group $P2_13$ (198). 
Several materials in this space group are chiral topological semimetals, such as RhSi, CoSi, AlPt, PdGa and PtGa\cite{Bradlyn2016,chang2017unconventional,sanchez2019topological,rao2019observation,schroter2019chiral,schroter2020observation,yao2020observation}.
The symmetry group has a primitive cubic Bravais lattice, and two additional symmetry generators which we take to be $\left\{C_{31}^+|000\right\}$ and $\left\{C_{2x}|\frac{1}{2}\frac{1}{2}0\right\}$. 
We shall include time reversal (TR) symmetry in the following discussion as well. 
According to Eq.~\eqref{eq:def_O}, for a rotationally-invariant set of basis orbitals that can be written in a spin-orbit decoupled basis (i.e. $s,p$ or $d$ orbitals) the (generally composite) band representation matrix for each symmetry generator $g$ is a tensor product of the Wyckoff position part, the spinless orbital part and the spin part
\begin{equation}\label{eq:tensorproductstructure}
\rho_{\bf k}(g) =\rho_{{\bf k},O}(g) \otimes \rho_{{\bf k},S}(g) \otimes\rho_{{\bf k},W}(g).
\end{equation}
Here $\rho_{{\bf k},O}(g)$ and $\rho_{{\bf k},S}(g)$ are the spinless and spinful representations for $g$ at the point $\mathbf{k}$, respectively. 
Note that this tensor product decomposition gives a slightly different basis for states than the canonical induced representation basis of Ref.~\cite{NaturePaper}.
We can determine the Wyckoff part by induction from the $s$-type orbitals at the $4a$ Wyckoff position, with coordinates
\begin{equation}
\left(\begin{array}{c}
\mathbf{q}_1 \\
\mathbf{q}_2 \\
\mathbf{q}_3 \\
\mathbf{q}_4
\end{array}\right)=\left(\begin{array}{ccc}
x & x & x \\
x+\half & \half-x & -x \\
-x & x+\half & \half-x \\
\half-x & -x & x+\half
\end{array}\right).
\end{equation}
To expose the role of multi-site orbital entanglement, we will start by taking as a basis for this band representation $s$ orbitals at each site $\mathbf{q}_i$. 
The site-symmetry group of the site $\mathbf{q}_1$ is generated by $C_{31}^+$, and is isomorphic to the cylcic group $C_3$. 
We give its character table in Appendix~\ref{sec:Some Useful Character Tables}. 
Using the formalism of Refs.~\cite{NaturePaper,cano2018building}, we can write the explicit form of the band representation matrices as
\begin{eqnarray}\begin{aligned}
\label{eq:bandrep_Wyckoff}
    \rho_{{\bf k},W}(C_{31}^+)&=\left(\begin{array}{cccc}
    1 & 0 & 0 & 0 \\
    0 & 0 & 0 & 1 \\
    0 & 1 & 0 & 0 \\
    0 & 0 & 1 & 0
    \end{array}\right),\\
    \rho_{{\bf k},W}(\{C_{2x}|\half\half 0\})&=\left(\begin{array}{cccc}
    0 & e^{ik_1} & 0 & 0 \\
    1 & 0 & 0 & 0 \\
    0 & 0 & 0 & e^{i(k_1-k_3)} \\
    0 & 0 & e^{-ik_3} & 0
    \end{array}\right),\\
    \rho_{{\bf k},W}(TR)&=i\left(\begin{array}{cccc}
    1 & 0 & 0 & 0 \\
    0 & 1 & 0 & 0 \\
    0 & 0 & 1 & 0 \\
    0 & 0 & 0 & 1
    \end{array}\right)\mathcal{K},
\end{aligned}\end{eqnarray}
in the basis of $(\mathbf{q}_1,\mathbf{q}_2,\mathbf{q}_3,\mathbf{q}_4)$. 
Here $\mathcal{K}$ is the complex conjugation operator. 
We are interested in the fourfold degeneracy at $\Gamma$, as well as the sixfold degeneracy at $R$. 
Let us focus first on the $\Gamma$ point. 

\subsection{$\Gamma$ point}
\label{sec:198_Gamma}

\subsubsection{s Orbitals}
Let us start by analyzing spinful $s$ orbitals at the $4a$ position, which induce the band representation $({}^1\bar{E}{}^2\bar{E}\uparrow G)_{4a}$. 
For the degeneracies at the $\Gamma$-point, it turns out that we can understand the physics much better with a change of basis. 
To see that, note that when $\mathbf{k}=0$ the vector $|\mathbf{q}_1\rangle+|\mathbf{q}_2\rangle+|\mathbf{q}_3\rangle+|\mathbf{q}_4\rangle$ is invariant under all of the matrices in Eq.~(\ref{eq:bandrep_Wyckoff}). 
This lets us block-diagonalize the $\rho_{{\bf k},W}(g)$, and we find that they split as
\begin{eqnarray}\begin{aligned}
\label{eq:split_wyckoff_gamma}
    \rho_{{\bf \Gamma},W}(C_{31}^+)&=1\oplus e^{-i\frac{2\pi}{3\sqrt{3}}(J^1_x+J^1_y+J^1_z)},\\
        \rho_{{\bf \Gamma},W}(\{C_{2x}|\half\half 0\})&=1\oplus e^{-i\pi J^1_x},\\
        \rho_{{\bf \Gamma},W}(\text{TR})&=i\mathcal{K}\oplus e^{-i\pi J^1_y}\mathcal{K}.\\        
\end{aligned}\end{eqnarray}

We notice that the first and second parts of the band representation are those for the spinless $s$ and $p$-orbitals respectively.
We have thus created linear combinations of orbitals from \emph{different sites}, which have the same orbital angular momentum as $s$ and $p$ orbitals placed at the origin of the unit cell in the \emph{symmorphic} space group $P23$. 
As a result, we shall label the states from the first direct summand as $|\bar{s}\rangle$ and those from the second direct summand as $|\bar{p}\rangle$. 
When we tensor product with spin $|\sigma\equiv\uparrow,\downarrow\rangle$ and subduce to the little group $G_\Gamma$, we will find that the $(|\bar{s}\uparrow\rangle, |\bar{s}\downarrow\rangle)$ state span a spin-momentum locked Kramers-Weyl fermion with $\mathbf{k}\cdot\mathbf{p}$ Hamiltonian 
\begin{equation}
H\propto \delta\mathbf{k}\cdot\mathbf{J}^{1/2},
\end{equation}
and sRDM determined by
\begin{equation}
 \mathbf{n}_{{1/2}}^{\pm1/2}(\delta\mathbf{k})=\hat{\mathbf{{\delta k}}},
\end{equation}
  as given in Eq.~\eqref{eq:spinRDM_s}.

Similarly, the states built out of $|\bar{p}\rangle\otimes|\sigma\rangle$ will behave like $P_{1/2}$ and $P_{3/2}$ orbitals at the origin of the unit cell, with sRDM given in Table.~\ref{tab:207} (note, however, that since space group $198$ has fewer symmetry constraints than space group $207$, the sRDMs of the two will generically differ when the Hamiltonian is perturbed away from the exactly solvable limit\cite{chang2017unconventional,flicker2018chiral}). 
Thus, the spin-momentum locking at $\Gamma$ for $s$ orbitals---where the pseudospin space is the physical spin space---is entirely captured by ``effective" orbitals located at the origin of the unit cell. 
We summarize the results in the $s$-orbital column of Table~\ref{tab:198_Gamma}.

Note that unlike in space group 207, the two twofold degeneracies subduced by the $s$ orbitals at the $\Gamma$ point are actually type-I entangled. 
This is because the twofold degeneracy from $\ket{\bar{s}}\otimes\ket{\sigma}$ and the twofold degeneracy from $\ket{\bar{p}}\otimes\ket{\sigma}$ are both isomorphic to the same two-dimensional irrep of $G_\Gamma$. 
In principle this means that the basis orbitals for each copy of this degeneracy are not uniquely determined, per our discussion in Sec.~\ref{sec:Composite band representations}. 
However, here we have chosen a physically-motivated way to specify the basis, by focusing on the Wyckoff position transformation and introducing effective $\ket{\bar{s}}$ and $\ket{\bar{p}}$ orbitals. 
This corresponds to treating band splitting due to intra-unit cell hopping as dominant over longer range hopping terms. 
We will continue with this choice throughout our analysis of $p$ and $d$ orbitals. 
One should keep in mind, however, that perturbations away from this limit will change the basis for the type-I entangled degeneracies of the same type, and hence perturb the entries in Table~\ref{tab:198_Gamma}.

\subsubsection{p Orbitals}
The sRDM at $\Gamma$ from $p$-orbitals can be similarly classified using the same ``trick'' of effective orbitals composed of different Wyckoff sites. 
The results will depend on whether on-site SOC or crystal field splitting is the dominant energy scale, as we have seen in Sec.~\ref{sec:Composite band representations}. 
Let us first suppose that on-site SOC dominates, and the five spinful $p$-orbitals first split into $P_{1/2}$ and $P_{3/2}$ orbitals. 
The $P_{1/2}$ orbitals transform in the ${}^1\bar{E}{}^2\bar{E}$ representation of the site-symmetry group, just like the $S_{1/2}$ orbitals (see Appendix~\ref{sec:Some Useful Character Tables} and Table~\ref{tab:c3chars}). 
The $P_{3/2}$ states, on the other hand, further split into the linear combination ${}^1\bar{E}{}^2\bar{E}\oplus\bar{E}\bar{E}$ of site symmetry group representations, where the first summand corresponds to the $\ket{J=3/2,m=\pm1/2}$ states, and the second summand corresponds to the $\ket{J=3/2,m=\pm3/2}$, where the spin quantization axis is taken along the $C_3$ axis $(111)$. 
The Clebsch-Gordan coefficients allow us to express each of these states in the spin and orbital basis, as we have done in the previous section. 
Note that the band representations induced by the two ${}^1\bar{E}{}^2\bar{E}$ site symmetry group representations are generically type-I entangled.

Once we have the states that carry the irreps of the site-symmetry group, we can proceed to consider the Wyckoff part of the symmetry transformation as we did for $s$ orbitals above. 
This is easiest in the basis of $P_{1/2}\oplus P_{3/2}$, where since the basis states are rotationally invariant we have---as a refinement of Eq.~\eqref{eq:tensorproductstructure} that the band representation matrices factorize as
\begin{equation}\label{eq:tensorproduct2}
\rho_\mathbf{k}(g) = \rho_{\mathbf{k},SO}(g)\otimes \rho_{\mathbf{k},W}(g).
\end{equation}
In words, the Wyckoff part of the symmetry transformation is simply an additional tensor product factor on the spin-orbital part. 

We then use the same effective $\ket{\bar{s}}$ and $\ket{\bar{p}}$ bases for the Wyckoff part. 
Since $\ket{\bar{s}}$ is a single state, $P_{1/2}\otimes\ket{\bar{s}}$ and $P_{3/2}\otimes\ket{\bar{s}}$ states will each contribute to the sRDM as if they are placed at the unit cell center. 
Thus, the twofold-degenerate representations at $\Gamma$ coming from from the ${}^1\bar{E}{}^2\bar{E}\otimes\ket{\tilde{s}}$ and the $\bar{E}\bar{E}\otimes\ket{\tilde{s}}$ representations will have the same $\mathbf{k}\cdot\mathbf{\sigma}$ invariant Hamiltonian as the twofold degeneracy at $\Gamma$ in space group $207$ (up to complex conjugation). 
The spin texture will thus be the same as listed in Table.~\ref{tab:207}, although $|\mathbf{n}(\delta{\bf k})|$ depends sensitively on the Clebsch-Gordan coefficients.
However, the situation for the space spanned by $|\bar{p}\rangle$ is a bit more complicated. 
Recall that  $|\bar{p}\rangle$ behaves effectively as spinless $p$-orbital with total angular momentum $j=1$. 
Upon taking a tensor product with the $P_{1/2} \oplus P_{3/2}$ states 
using the language of addition of angular momentum\cite{griffiths2018introduction}, we can keep track of the dimension for the Hilbert space spanned by $|\bar{p}\rangle$ as
\begin{eqnarray}\begin{aligned}
\label{eq:198_Gamma_p}
{\bf 3}\otimes({\bf 4}\oplus{\bf 2}) &= ({\bf 6}\oplus{\bf 4}\oplus{\bf 2})\oplus({\bf 4}\oplus{\bf 2})\\
&=(({\bf 4}\oplus{\bf 2})\oplus{\bf 4}\oplus{\bf 2})\oplus({\bf 4}\oplus{\bf 2}).
\end{aligned}\end{eqnarray}
Here the factors at the left hand side indicates the dimension of $|\bar{p}\rangle$ and the $P_{3/2}\oplus P_{1/2}$ orbitals respectively. 
The first equality results from the effective SOC between these degrees of freedom.
In the last line of Eq.~\eqref{eq:198_Gamma_p}, the states with total angular momentum $j=5/2$, denoted by ${\bf 6}$, further split into a fourfold degeneracy and a twofold degeneracy due to the crystal field, as discussed in Appendix~\ref{sec:Some Useful Character Tables}.
Thus, we see that with high multiplicity Wyckoff positions, the type-I entanglement of $p$ orbitalS can exhibit a rich structure. 
A repeated application of the Clebsch-Gordan decomposition then yields the sRDMs summarized in the rows labelled by ``SOC'' in Table.~\ref{tab:198_Gamma}. 
We find that the sRDMs for the $P_{1/2}$ orbitals give radial spin texture consistent with what we see for $s$ orbitals, although with the vector $\mathbf{n}$ divided by $3$ due to the Clebsch-Gordan coefficients. 
For the $P_{3/2}$ orbitals we find spin textures more reminiscent of those for $d$ orbitals in space group 207 (Table~\ref{tab:207}). 
We can understand this through addition of angular momentum: $P_{3/2}\otimes\ket{\bar{s}}$ states behave like an effective spin-$3/2$, while $P_{3/2}\otimes\ket{\bar{p}}$ behaves like an effective $D_{5/2}$ state.

The situation is very different when the crystal field splitting dominates over on-site SOC. 
In this case, the spinless $p$-orbitals at the $4a$ position first split into two representations of the site symmetry group: a one-dimensional representation $A$ spanned by the $p_x+p_y+p_z$ orbital, and a two-dimensional representation ${}^1E{}^2E$ spanned by the $(p_x-2p_y+p_z, p_x-p_z)$ orbitals (see Appendix~\ref{sec:Some Useful Character Tables} for further details). 
The $p_x+p_y+p_z$ orbital will behave like a spinless $s$-orbital when coupled to the Wyckoff positions because the orbitals at different sites are related by symmetry. 
To see this explicitly, we can introduce the ``dressed'' Wyckoff positions as follows
\begin{eqnarray}
\begin{aligned}
\label{eq:dress_wyckoff}
 |\bar{\bf q}_1\rangle &\equiv |{\bf q}_1\rangle\otimes\frac{|p_x\rangle+|p_y\rangle+|p_z\rangle}{\sqrt{3}},  \\
 |\bar{\bf q}_2\rangle &\equiv |{\bf q}_2\rangle\otimes\frac{|p_x\rangle-|p_y\rangle-|p_z\rangle}{\sqrt{3}}
 =\rho_{{\bf \Gamma},W\otimes O}(C_{2x})|\bar{\bf q}_1\rangle, \\
 |\bar{\bf q}_3\rangle &\equiv |{\bf q}_3\rangle\otimes\frac{-|p_x\rangle+|p_y\rangle-|p_z\rangle}{\sqrt{3}} 
 =\rho_{{\bf \Gamma},W\otimes O}(C_{2y})|\bar{\bf q}_1\rangle, \\
 |\bar{\bf q}_4\rangle &\equiv |{\bf q}_4\rangle\otimes\frac{-|p_x\rangle-|p_y\rangle+|p_z\rangle}{\sqrt{3}}
 =\rho_{{\bf \Gamma},W\otimes O}(C_{2z})|\bar{\bf q}_1\rangle,
\end{aligned}
\end{eqnarray}
where $\rho_{{\bf \Gamma},W\otimes O}(g)\equiv\rho_{{\bf \Gamma},W}(g)\otimes\rho_{{\bf \Gamma},O}(g)$ is the orbital and Wyckoff parts of the representation at $\Gamma$, and $C_{2x}$, $C_{2y}=C_{31}^+C_{2x}(C_{31}^+)^{-1}$, $C_{2z}=C_{31}^+C_{2y}(C_{31}^+)^{-1}$. 
This corresponds to the basis that would be obtained from the canonical induction formula.
It is straightforward to confirm that the band representation matrices in this space are exactly the same as that in Eq.~\eqref{eq:bandrep_Wyckoff}, which are those for spinless $s$-orbitals. 
Hence we expect that the sRDM of the $p_x+p_y+p_z$ orbital will be exactly the same as that of the $s$-orbital. 
For the space spanned by the $(p_x-2p_y+p_z, p_x-p_z)$ orbitals, we can similarly introduce the ``dressed'' Wyckoff positions in this space, followed by splitting them into effective $|\bar{s}\rangle$ and $|\bar{p}\rangle$ orbitals. 
It turns out that there are two twofold and three fourfold degeneracies subduced at $\Gamma$ in this space (see App.~\ref{sec:bandrep_p_crystal}), corresponding to the $[({}^1E{}^2E\otimes {}^1\bar{E}{}^2\bar{E}) \uparrow G]_{4a} = (\bar{E}\bar{E}\uparrow G)_{4a}\oplus ({}^1\bar{E}{}^2\bar{E} \uparrow G)_{4a}$ composite band representation. 
Combining the formulas for the basis states with the $\delta\mathbf{k}\cdot\mathbf{J}$ form of the $\mathbf{k}\cdot\mathbf{p}$ Hamiltonian, we can compute the sRDM as in the previous sections.
The sRDMs are tabulated in the rows labelled by ``Crystal field'' in Table.~\ref{tab:198_Gamma}. 
The sRDM we find here include fourfold degeneracies with spin winding numbers $\pm 5$ and $\pm 1$, as well as twofold degeneracies with radial spin texture.

\subsubsection{d Orbitals}

The sRDM of the $d$-orbitals can be calculated similarly. 
Let us first consider the case when on-site SOC is dominant. 
In this case, the $d$-orbitals on each site first split into  $D_{3/2}$ and $D_{5/2}$ orbitals. $D_{3/2}$ orbitals transform under the site-symmetry group identically to $P_{3/2}$ orbitals above.

\begin{figure}[ht]
\includegraphics[width=\columnwidth]{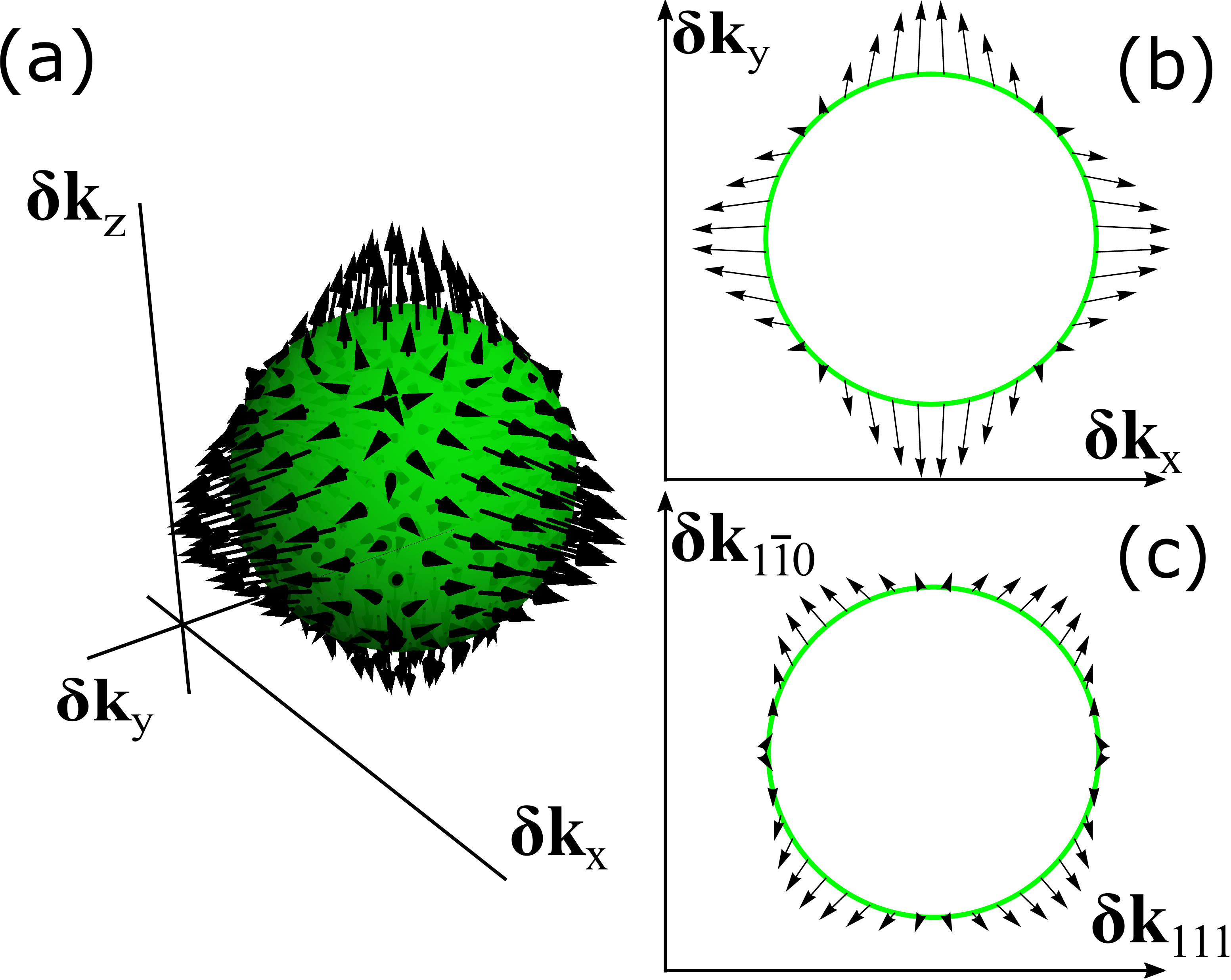}
\caption{
Plot of the spin texture with winding numbers $\nu=-7$ for crystal-field split $d$-orbitals in space group $P432$ (207). a) shows the vector $\mathbf{n}(\mathbf{k})$ on a constant-energy sphere. b) shows the projection $(n_x(\mathbf{k}),n_y(\mathbf{k}))$ of the spin texture along the projection of the constatnt energy sphere to the $k_x-k_y$ plane. c) shows the analogous projection in the $k_{111}-k_{1\bar{1}0}$ plane.
}\label{fig:winding_7}
\end{figure}

For the $D_{5/2}$ orbitals, we again use the fact that for an $SU(2)$-invariant set of orbitals the band representation matrices have a tensor product structure Eq.~\eqref{eq:tensorproduct2}. 
This lets us focus on the Wyckoff and spin-orbital part of the band representation matrices separately. 
Focusing on the spin-orbital part, we have at the $\Gamma$ point that the band representation matrices are
\begin{align*}
    \rho^{5/2}_{\bf \Gamma}(C_{31}^+) &= e^{-i\frac{2\pi}{3\sqrt{3}}(J_x^{3/2}+J_y^{3/2}+J_z^{3/2})}\oplus e^{-i\frac{2\pi}{3\sqrt{3}}(J_x^{1/2}+J_y^{1/2}+J_z^{1/2})}\\
    \rho^{5/2}_{\bf \Gamma}(C_{2x}) &= e^{-i\pi J_x^{3/2}}\oplus e^{-i\pi J_x^{1/2}}\\
    \rho^{5/2}_{\bf \Gamma}(TR) &= e^{-i\pi J_y^{3/2}}\oplus e^{-i\pi J_y^{1/2}}
\end{align*}
which coincide with the band representations for $P_{3/2}$ and $P_{1/2}$ orbitals at $\Gamma$, respectively. 
However, this does not imply that the $d$-orbital have the same sRDMs as the $p$-orbital, because the sRDM also depends on the basis functions (Clebsch-Gordan coefficients) as we have seen in Sec.~\ref{sec:Composite band representations}. 
Nevertheless, the arguments presented above for $p$-orbitals also apply for the $d$-orbitals if we substitute the Clebsch-Gordan coefficients of the latter to obtain the spin-orbit decoupled states. 
The results are presented in Table.~\ref{tab:198_Gamma}. 
Interestingly, in addition to bands with spin winding number $\pm 1$ and $\pm 5$, we find that with $D_{5/2}$ orbitals we can get an sRDM at $\Gamma$ for a fourfold degeneracy with spin winding numbers
\begin{align}
\nu_{3/2}^{\pm 3/2}&=\mp 1, \\
\nu_{3/2}^{\pm 1/2}&=\pm 7.
\end{align}
We can understand this as originating from the $D_{5/2}\otimes\ket{\bar{p}}$ states, which include an effective spin-$7/2$ orbital. In Fig.~\ref{fig:winding_7} we show a plot of the spin texture with winding $-7$.

The sRDMs of the $d$-orbitals are in fact given for free when the crystal field is dominant. 
To obtain a basis for the crystal-field split orbitals at each site, we can use the fact that the site symmetry group $C_3$ (and in fact the entire little group $G_\Gamma$) is contained in the chiral octahedral group discussed in Sec.~\ref{sec:1apos}. 
Recall from Sec.~\ref{sec:Type-I entanglement: degeneracies with non-minimal orbital content} that 
if the crystal field splitting dominates over the on-site SOC, the spinless $d$-orbitals split into a three-dimensional representation of the chiral octahedral group carried by the $(d_{xy},d_{xz},d_{yz})$ orbitals, and a two-dimensional representation carried by the $(d_{x^2-y^2},d_{z^2})$ orbitals. 
If we now restrict to the point group of space group $P2_13 $ (198), we see in Table.~\ref{tab:207_p_d_symmetry} that the $(d_{xy},d_{xz},d_{yz})$ transform exactly like $p$-orbitals, and hence give the same sRDMs. 
For the $(d_{x^2-y^2},d_{z^2})$ orbitals, it turns out that they carry the same band representation as the $(p_x-2p_y+p_z, p_x-p_z)$ orbitals, since they both carry the ${}^1E{}^2E$ representation of the site symmetry group. 
In contrast to the strong-SOC limit, these band representations have isomorphic spin-orbit decoupled bases, and hence the sRDMs induced from $(d_{x^2-y^2},d_{z^2})$ orbitals coincide with those listed in the rows labelled by $p_x-2p_y+p_z, p_x-p_z$ in Table.~\ref{tab:198_Gamma}.

\begin{widetext}

\begin{table}[htp]
\begin{center}
\begin{tabular}{|c|c|c|c|c|c|c|c|}
\hline
\multicolumn{3}{|c|}{$s$-orbital} & \multicolumn{5}{|c|}{$p$-orbital}  \\
\hline
\multicolumn{2}{|c|}{} & & & & \multicolumn{2}{|c|}{twofold}  & $\mp\frac{1}{3}\delta{\bf k}~[\mp1]$, $\pm\frac{1}{9}\delta{\bf k}~[\pm1]$ \\
\cline{6-8}
\multicolumn{2}{|c|}{} & & & $P_{1/2}$ & \multirow{2}{*}[-0.05cm]{fourfold} & $m=\pm1/2$ & $\mp\frac{1}{9}\delta{\bf k}~[\mp1]$ \\ 
\cline{7-8}
\multicolumn{2}{|c|}{twofold}  &$\pm\delta{\bf k}~[\pm1]$  & \parbox[t]{2mm}{\multirow{3}{*}{\rotatebox[origin=c]{90}{SOC}}} &  &  & $m=\pm3/2$ & $\mp\frac{1}{3}\delta{\bf k}~[\mp1]$ \\
\cline{5-8}
\multicolumn{2}{|c|}{} &$\mp\frac{1}{3}\delta{\bf k}~[\mp1]$ & & & \multicolumn{2}{|c|}{twofold}  & $\pm\frac{5}{9}\delta{\bf k}~[\pm1]$, $\pm\frac{1}{3}\delta{\bf k}~[\pm1]^*$ \\
\cline{6-8}
\multicolumn{2}{|c|}{} & & & $P_{3/2}$ & \multirow{2}{*}[-0.05cm]{fourfold} & $m=\pm1/2$ & $\pm\frac{1}{3}\delta{\bf k}~[\pm1]$, $\pm\frac{11}{45}\delta{\bf k}~[\pm1]$, $\mp(\frac{16}{15}\delta{\bf k}-\frac{9}{5}\delta{\bf k}^3)~[\pm5]$ \\
\cline{7-8}
\multicolumn{2}{|c|}{} & &  &  &  & $m=\pm3/2$ & $\pm\delta{\bf k}~[\pm1]$, $\pm\frac{11}{15}\delta{\bf k}~[\pm1]$, $\pm(\frac{2}{5}\delta{\bf k}-\frac{3}{5}\delta{\bf k}^3)~[\mp5]$ \\
\hline
& & & & & \multicolumn{2}{|c|}{twofold}  & $\pm\delta{\bf k}~[\pm1]$, $\mp\frac{1}{3}\delta{\bf k}~[\mp1]$ \\
\cline{6-8}
& $m=\pm1/2$ & $\pm\frac{1}{3}\delta{\bf k}~[\pm1]$ & & $p_x+p_y+p_z$ & \multirow{2}{*}[-0.05cm]{fourfold} & $m=\pm1/2$ &  $\pm\frac{1}{3}\delta{\bf k}~[\pm1]$ \\ 
\cline{7-8}
\multirow{2}{*}[-0.05cm]{fourfold}&  &  & \parbox[t]{2mm}{\multirow{3}{*}[0.3cm]{\rotatebox[origin=c]{90}{Crystal field}}} &  &  & $m=\pm3/2$ & $\pm\delta{\bf k}~[\pm1]$\\
\cline{2-3}\cline{5-8}
&  & & & & \multicolumn{2}{|c|}{twofold}  & $\mp\frac{1}{3}\delta{\bf k}~[\mp1]$, $\mp\frac{1}{3}\delta{\bf k}~[\mp1]$ \\
\cline{6-8}
& $m=\pm3/2$ & $\pm\delta{\bf k}~[\pm1]$ & & 
$\begin{array} {lcl} &p_x-2p_y+p_z& \\ &p_x-p_z& \end{array}$ 
& \multirow{2}{*}[-0.05cm]{fourfold} & $m=\pm1/2$ &
$\begin{array} {lcl} 
&\pm(2\delta{\bf k}-3\delta{\bf k}^3)~[\mp5]^*, \mp(\frac{2}{3}\delta{\bf k}-\delta{\bf k}^3)~[\pm5]^*& \\ 
&\pm(\frac{2}{3}\delta{\bf k}-\delta{\bf k}^3)~[\mp5]&
\end{array}$
\\
\cline{7-8}
& & &  &  &  & $m=\pm3/2$ & $\pm\delta{\bf k}^3~[\pm1]^*$, $\mp\frac{1}{3}\delta{\bf k}^3~[\mp1]^*$, $\pm\frac{1}{3}\delta{\bf k}^3~[\pm1]$ \\
\hline
\multicolumn{8}{|c|}{$d$-orbital}\\
\hline
& & \multicolumn{2}{|c|}{twofold} & \multicolumn{4}{|c|}{$\mp\frac{1}{3}\delta{\bf k}~[\mp1]$, $\mp\frac{1}{5}\delta{\bf k}~[\mp1]^*$}\\
\cline{3-8}
& $D_{3/2}$ & \multirow{2}{*}[-0.05cm]{fourfold}& $m=\pm1/2$ & \multicolumn{4}{|c|}{$\mp\frac{1}{5}\delta{\bf k}~[\mp1]$, $\mp\frac{11}{75}\delta{\bf k}~[\mp1]$, $\pm(\frac{16}{25}\delta{\bf k}-\frac{27}{25}\delta{\bf k}^3)~[\mp5]$} \\
\cline{4-8}
\parbox[t]{2mm}{\multirow{3}{*}{\rotatebox[origin=c]{90}{SOC}}}& & & $m=\pm3/2$ & \multicolumn{4}{|c|}{$\mp\frac{3}{5}\delta{\bf k}~ [\mp1]$, $\mp\frac{11}{25}\delta{\bf k}~[\mp1]$, $\mp(\frac{6}{25}\delta{\bf k}-\frac{9}{25}\delta{\bf k}^3)~[\pm5]$} \\
\cline{2-8}
& & \multicolumn{2}{|c|}{twofold} & \multicolumn{4}{|c|}{$\pm\frac{1}{3}\delta{\bf k}~ [\pm1]^*$, $\mp\frac{1}{9}\delta{\bf k}~[\mp1]^*$, $\pm\frac{1}{45}\delta{\bf k}~[\pm1]$, $\pm\frac{1}{3}\delta{\bf k}~[\pm1]^*$}\\
\cline{3-8}
& $D_{5/2}$ & \multirow{2}{*}[-0.05cm]{fourfold}& $m=\pm1/2$ & \multicolumn{4}{|c|}{$\mp(\frac{16}{15}\delta{\bf k}-\frac{27}{15}\delta{\bf k}^3)~[\pm5]$, $\pm\frac{1}{9}\delta{\bf k}~[\pm1]^*$, $\pm(\frac{148}{225}\delta{\bf k}-\frac{27}{25}\delta{\bf k}^3)~[\mp5]$, $\pm(\frac{2}{15}\delta{\bf k}-\frac{9}{25}\delta{\bf k}^3)~[\pm7]$} \\
\cline{4-8}
& & & $m=\pm3/2$ & \multicolumn{4}{|c|}{$\pm(\frac{2}{5}\delta{\bf k}-\frac{3}{5}\delta{\bf k}^3)~[\mp5]$, $\pm\frac{1}{3}\delta{\bf k}~[\pm1]^*$, $\mp(\frac{14}{75}\delta{\bf k}-\frac{9}{25}\delta{\bf k}^3)~[\pm5]$, $\mp(\frac{8}{25}\delta{\bf k}-\frac{3}{25}\delta{\bf k}^3)~[\mp1]$}\\
\hline
\end{tabular}
\end{center}
\caption{
The sRDMs for space group $P2_13 $ (198) at $\Gamma$-point. 
Entries separated by commas correspond to different degeneracy points subduced by the same set of orbitals. 
For orbitals that subduce multiple fourfold degeneracies, the $m=\pm1/2$ and $m=\pm 3/2$ for each degeneracy appear in the same order in both rows. 
Here $\hat{\bf \delta k}^3$ should be understood as a vector $(\hat{\delta k}_x^3,\hat{\delta k}_y^3,\hat{\delta k}_z^3)$. 
The square brakets indicate the corresponding winding numbers of the degeneracies.
An asterisk is used to denote cases where the invariant Hamiltonian has the form $\delta\mathbf{k}\cdot\mathbf{J}^*$.
When crystal field splitting dominates, the sRDMs induced from the $d$-orbitals can be determined from the corresponding rows for the $p$-orbitals, as described in the main text. 
}
\label{tab:198_Gamma}
\end{table}%

\end{widetext}

\begin{table}[htp]
\begin{center}
\begin{tabular}{|c|c|c|c|c|c|}
\hline
& $\left\{C_3|000\right\}$ & $\left\{C_{2x}|\frac{1}{2}\frac{1}{2}0\right\}$ \\
\hline
$p_{x,y,z}$ & $\begin{pmatrix}
& 1&  \\
 & & 1\\
1& 
\end{pmatrix}$
& 
$\begin{pmatrix}
1& &  \\
 & -1& \\
& & -1
\end{pmatrix}$
 \\
 \hline
$d_{xy},d_{yz},d_{zx}$ & $\begin{pmatrix}
& 1&  \\
 & & 1\\
1& 
\end{pmatrix}$
& 
$\begin{pmatrix}
1& &  \\
 & -1& \\
& & -1
\end{pmatrix}$
 \\
  \hline
$d_{x^2-y^2},d_{z^2}$ & $\begin{pmatrix}
-\frac{1}{2} & -\frac{\sqrt{3}}{2} \\
\frac{\sqrt{3}}{2} & -\frac{1}{2}
\end{pmatrix}$
& 
$\begin{pmatrix}
1 \\
& 1
\end{pmatrix}$
\\
\hline
\end{tabular}
\end{center}
\caption{The representations for the symmetry generators in space group $P2_13 $ (198) for different spinless orbitals. }
\label{tab:207_p_d_symmetry}
\end{table}%

\subsection{$R$ Point}
\label{sec:198_R}

Let us now move on to discuss the sRDM at the $R$ point. 
Unlike the situation for $\Gamma$-point, there is no simple basis to diagonalize the band representation matrices for the Wyckoff part alone. 
To analyze the sRDM in this case, we will again consider $s$, $p$, and $d$ orbitals separately.

\subsubsection{s Orbitals}\label{sec:198_R_s}
Since spinless $s$ orbitals are scalars, we need only consider $\rho_{{\bf k},S\otimes W}(g) \equiv\rho_{{\bf k},S}(g)\otimes\rho_{{\bf k},W}(g) $, the tensor product of the spin and Wyckoff representation matrices for the symmetry elements. 
Upon substituting the coordinates of the $R$ point ${\bf k}=(\pi,\pi,\pi)$ into Eq.~\eqref{eq:bandrep_Wyckoff}, we can block-diagonalize this representation. 
We find that there is a two-dimensional invariant subspace spanned by the vectors
\begin{align}
\label{eq:rpoint2foldspace}
    |+\rangle&=\frac{1}{2}\left(i|\mathbf{q}_1\downarrow\rangle - |\mathbf{q}_2\uparrow\rangle + i|\mathbf{q}_3\uparrow\rangle + |\mathbf{q}_4\downarrow\rangle\right),\nonumber \\
    |-\rangle&=\frac{1}{2}\left(-i|\mathbf{q}_1\uparrow\rangle + |\mathbf{q}_2\downarrow\rangle + i|\mathbf{q}_3\downarrow\rangle + |\mathbf{q}_4\uparrow\rangle\right),
\end{align}
as well as an orthogonal six-dimensional invariant subspace. 
A convenient basis for the orthogonal subspace is
\begin{eqnarray}
\label{eq:rpoint6foldspace}
\begin{aligned}
    |1a\rangle&=\frac{1}{2\sqrt{2}}(|\mathbf{q}_1\uparrow\rangle+|\mathbf{q}_2\uparrow\rangle+i|\mathbf{q}_3\uparrow\rangle-i|\mathbf{q}_4\uparrow\rangle\\
    &\quad\quad-i|\mathbf{q}_1\downarrow\rangle+i|\mathbf{q}_2\downarrow\rangle+|\mathbf{q}_3\downarrow\rangle+|\mathbf{q}_4\downarrow\rangle) \\
    |2a\rangle &= -\rho_{{R},S\otimes W}(C_{31}^+)|1a\rangle \\
    |3a\rangle & = \rho_{{R},S\otimes W}(C_{31}^+)^2|1a\rangle \\
    |1b\rangle&=\frac{1}{2\sqrt{2}}(i|\mathbf{q}_1\uparrow\rangle-i|\mathbf{q}_2\uparrow\rangle+|\mathbf{q}_3\uparrow\rangle+|\mathbf{q}_4\uparrow\rangle\\
    &\quad\quad-|\mathbf{q}_1\downarrow\rangle-|\mathbf{q}_2\downarrow\rangle+i|\mathbf{q}_3\downarrow\rangle-i|\mathbf{q}_4\downarrow\rangle) \\
    |2b\rangle &= -\rho_{{R},S\otimes W}(C_{31}^+)|1b\rangle \\
    |3b\rangle & = \rho_{{R},S\otimes W}(C_{31}^+)^2|1b\rangle.
\end{aligned}
\end{eqnarray}
Re-expressing the matrices $\rho_{R,S\otimes W}(g)$ in this basis, we find
\begin{eqnarray}\begin{aligned}
\label{eq:bandrep_198_R}
    \rho_{{R},S\otimes W}(C_3) &=e^{-i\frac{2\pi}{3\sqrt{3}}(-J^{1/2}_x+J^{1/2}_y-J^{1/2}_z)} \\
    &\quad\oplus \left(\mathbb{I}_2\otimes \begin{pmatrix}
& &1  \\
 1& & \\
& 1& 
\end{pmatrix}\right),\\
    \rho_{{R},S\otimes W}(\{C_{2x}|\half\half 0\})&= \mathbb{I}_2 \oplus  \left(\mathbb{I}_2\otimes \begin{pmatrix}
1& &  \\
 &-1 & \\
& &-1 
\end{pmatrix}\right),\\
    \rho_{{R},S\otimes W}(TR)&= ie^{i\pi J^{1/2}_y }\mathcal{K} \oplus  (ie^{-i\pi J^{1/2}_y}\otimes \mathbb{I}_3\mathcal{K} ),    
\end{aligned}\end{eqnarray}
where $\mathbb{I}_2$ is the $2\times2$ identity matrix. 

Examining the basis in Eq.~(\ref{eq:rpoint2foldspace}), we find that matrix elements for all spin operators vanish in this subspace. 
This means that regardless of the invariant Hamiltonian (in fact, the invariant Hamiltonian here is quadratic in $\mathbf{\delta k}$ to leading order), we have
\begin{eqnarray}\begin{aligned}
\label{eq:sRDM_R_2fold}
\mathbf{n} = {\bf 0}
\end{aligned}\end{eqnarray}
for the sRDM near the twofold degeneracy. 
In contrast to our previous examples, this shows that strongly spin-orbit coupled degeneracies can have trivial spin momentum locking. 
The lack of spin-momentum locking here originates from the high-degree of entanglement between spin and orbital degrees of freedom in Eq.~(\ref{eq:rpoint2foldspace}). 
This indicates that although the band representation is built from only $s$-orbitals, twofold degeneracies can range from perfectly spin-momentum locked to entirely mixed spin states. 

Next, we examine the sixfold degenerate subspace. 
We can write the most general ${\bf k}\cdot{\bf p}$ Hamiltonian in the six-fold degenerate subspace as\cite{flicker2018chiral}
\begin{eqnarray}
\begin{aligned}
\label{eq:6_fold_H}
    H_R({\bf \delta k},b,\varphi, v_f)&=\hbar v_f\left(
    \begin{array}{cc}
    H^{(3)}(\delta\mathbf{k},\varphi) & bH^{(3)}(\delta\mathbf{k},0) \\
    b^*H^{(3)}(\delta\mathbf{k},0) & -H^{(3)}(\delta\mathbf{k},-\varphi)
    \end{array}
    \right), \\
    H^{(3)}(\mathbf{\delta k},\varphi)&=\left(\begin{array}{ccc}
    0 & e^{i\varphi}k_z & e^{-i\varphi}k_y \\
    e^{-i\varphi}k_z & 0 & e^{i\varphi}k_x \\
    e^{i\varphi}k_y & e^{-i\varphi}k_x & 0
    \end{array}\right),
\end{aligned}
\end{eqnarray}
where $v_f$ is the Fermi velocity, and $b, \varphi$ are real-valued free parameters.
In the simplest case, we have $\varphi=\pi/2,b=0$, in which case we have a decoupled set of threefold degeneracies with $\delta\mathbf{k}\cdot\mathbf{J}$ Hamiltonians. 
In this limit we have that for each block, labeled as $\alpha=\pm$ respectively, the states behave exactly like a spinless $p$-orbital as evident from Eq.~\eqref{eq:bandrep_198_R}. 
Thus we shall label the states as $|\alpha=\pm,j=1,m\rangle$ to reflect that the state has total angular momentum $j=1$. 
With that, we have the eigenstate of the Hamiltonian $H_R({\bf \delta k},b=0,\varphi=\pi/2,v_f)$ are
\begin{eqnarray}\begin{aligned}
\label{eq:198_eigvec_1}
|\alpha=\pm,\delta{\bf k}m\rangle = e^{-i\phi J^1_z}e^{-i\theta J^1_y}|\alpha=\pm,j=1,m\rangle.
\end{aligned}\end{eqnarray}
Upon decomposing the eigenstates to spin-orbit decoupled states using Eq.~\eqref{eq:rpoint6foldspace}, the sRDM is calculated to be
\begin{eqnarray}\begin{aligned}
\label{eq:spinRDM_198_R_6fold}
n^{\alpha,m}_{1,i} &=\frac{m}{2}\hat{\delta k}_i +\alpha \frac{2-3|m|}{2}\lambda_{ijk}\hat{\delta k}_j\hat{\delta k}_k,
\end{aligned}\end{eqnarray}
where $\lambda_{ijk}$ is a fully symmetric tensor equal to $1/2$ whenever $(ijk)$ is a permutation of $(123)$, and zero otherwise (in other words, it is half the absolute value of the Levi-Civita tensor). 
We recall that ${\bf n}^{\alpha,m}_j$ are the sRDM for the state $|\alpha,j,m\rangle$, 
and unlike the isotropic spin-momentum locking for twofold and fourfold degenerate fermions, here the spin-momentum locking explicitly reflects the threefold cubic symmetry.
\begin{figure}[t]
\includegraphics[width=\columnwidth]{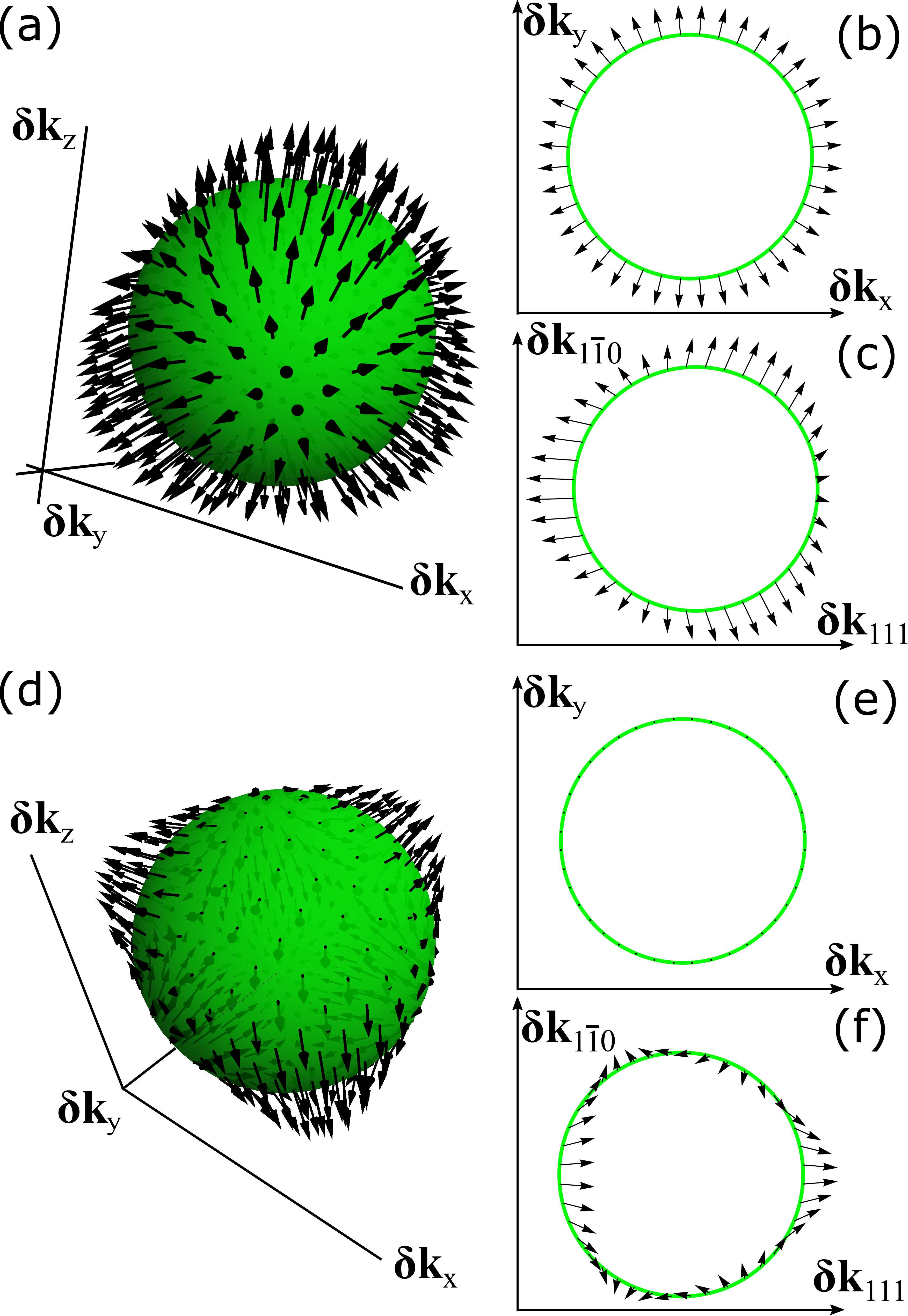}
\caption{Representation of the spin texture for the sixfold degeneracy at the $R$ point in space group $P2_13$ (198) for $|m|=1$ (a,b,c) and $|m|=0$ (d,e,f). a) shows the vector $\mathbf{n}(\mathbf{k})$ on a constant-energy sphere for $|m|=1$. b) shows the projection $(n_x(\mathbf{k}),n_y(\mathbf{k}))$ of the spin texture along the projection of the constatnt energy sphere to the $k_x-k_y$ plane for $|m|=1$. c) shows the analogous projection in the $k_{111}-k_{1\bar{1}0}$ plane for $|m|=1$. d) shows the vector $\mathbf{n}(\mathbf{k})$ on a constant-energy sphere for $|m|=0$. e) shows the projection $(n_x(\mathbf{k}),n_y(\mathbf{k}))$ of the spin texture along the projection of the constatnt energy sphere to the $k_x-k_y$ plane for $|m|=0$. f) shows the analogous projection in the $k_{111}-k_{1\bar{1}0}$ plane for $|m|=0$.}\label{fig:texture}
\end{figure}
In Fig.~\ref{fig:texture}, we show the representations of this spin texture for $\alpha=+$ and $|m|=0,1$.
With Eq.~\eqref{eq:chern_number}, we find that the winding numbers of the $\mathbf{n}$ vector for the $|\alpha=\pm,\delta{\bf k}m\rangle$ states read
\begin{align}
\nu_{j=1}^{\alpha,m}&=m,
\end{align}
which is exactly the magnetic quantum number of the bands. 
Since the winding number is quantized, we expect that it will not change when the Hamiltonian is weakly perturbed away from the exactly solvable point, as discussed in Sec.~\ref{sec:1apos}. 
Furthermore, it should be robust to the addition of more band representations into the spectrum, provided $|\nu_1^{\alpha,m}|\neq 0$.
It is important to note that although the bands with different values of $\alpha$ feature distinct patterns of spin-momentum locking, they are in fact degenerate when $b=0,\phi=\pi/2$ (or when $b$ satisfies certain condition, see Eq.~\eqref{eq:relation_b_main}). 
From an experimental point of view, this implies that a probe that is sensitive only to spin will measure the sRDM averaged over different $\alpha$ sectors in this limit, thus reproducing a radial spin texture. 
However, for generic values of $b$ and $\phi$, which are material-dependent, this degeneracy could be lifted along most directions, allowing for the sRDM of bands with different $\alpha$ to be distinguished. 
In that case, one could fit the measured spectrum against the model In Eq.~\eqref{eq:6_fold_H}, followed by measuring the sRDMs in different $\alpha$-sectors.

\subsubsection{p Orbitals}

The sRDM of the $p$-orbitals will, again, depend on the relative strength of the on-site SOC and crystal field splitting. 
Let us consider first the case where the spinful $p$-orbital is split into $P_{1/2}$ and $P_{3/2}$ orbitals due to SOC. 
The band representation induced from $P_{1/2}$ orbitals subduces a twofold and a sixfold degeneracies, in a similar basis as in Eqs.~\eqref{eq:rpoint2foldspace}-\eqref{eq:rpoint6foldspace} for $s$ orbitals, with the spin state $|\uparrow\rangle$ ($|\downarrow\rangle$) replaced by spin-orbit coupled state $|j=1/2, m=1/2\rangle$ ($|j=1/2, m=-1/2\rangle$). 
The resulting sRDMs are those in Eqs.~\eqref{eq:sRDM_R_2fold} and \ref{eq:spinRDM_198_R_6fold} multiplied by a factor of $-1/3$, an explicit manifestation of the Clebsch–Gordan coefficients. 
We can follow a similar construction for $P_{3/2}$ orbitals, and the resulting sRDMs are listed in the rows of Table.~\ref{tab:198_R} labelled by ``SOC''. 
Again we find that all twofold degeneracies at the $R$ point have vanishing $\mathbf{n}$. 
For the sixfold degeneracies, we find that one of the sixfold degeneracies subduced from $P_{3/2}$ orbitals has spin winding $\nu_1^{\alpha m} = 3 m$.

When crystal field splitting dominates, we again find that the space spanned by the $p_x+p_y+p_z$ orbital behaves like that spanned by spinless $s$-orbitals, after coupling to the Wyckoff position part. 
To see that, we can substitute the ``dressed'' Wyckoff positions (see Eq.~\eqref{eq:dress_wyckoff}) into the basis in Eqs.~\eqref{eq:rpoint2foldspace}--\eqref{eq:rpoint6foldspace}, and show that the band representation matrices reduce to those in Eq.~\eqref{eq:bandrep_198_R} for $s$-orbitals. 
Hence, the sRDMs from the $p_x+p_y+p_z$ orbital are exactly those in Eqs.~\eqref{eq:sRDM_R_2fold} and \eqref{eq:spinRDM_198_R_6fold}. 
For the space spanned by the $(p_x-2p_y+p_z, p_x-p_z)$ orbitals, with a similar substitution of the ``dressed'' Wyckoff positions (see App.~\ref{sec:bandrep_p_crystal}) we can show that the band representations split into two copies of twofold and sixfold degeneracies respectively, with the sRDMs shown in the rows labelled by ``Crystal field'' in Table.~\ref{tab:198_R}. 
These are similar in structure to the sRDMs we found for $P_{3/2}$ orbitals.

\subsubsection{d Orbitals}

For the sRDM from $d$-orbitals, our discussion at the end of Sec.~\ref{sec:198_Gamma} still applies. 
As evident from Eq.~\eqref{eq:bandrep_Wyckoff}, the only difference between the band representation matrices at the $\Gamma$ and $R$-points lies in the representation $\rho_{{\bf k},W}(\{C_{2x}|\half\half 0\})$. 
This, however, will not change the fact that spinful $d$-orbitals, after splitting by strong on-site SOC, induce the same band representations as $p$-orbitals.
As a result, we can follow the same calculation as for $p$ orbitals, followed by substituting the Clebsch-Gordan coefficients for $d$-orbitals to obtain the sRDMs. 
These are presented at the bottom of Table~\ref{tab:198_R}.
Similarly, note that the orbital part of the symmetry transformations for $(d_{xy}, d_{yz}, d_{zx})$ and $(d_{x^2-y^2}, d_{z^2})$ subspaces, as presented in Table.~\ref{tab:207_p_d_symmetry} are independent of $\mathbf{k}$. 
Thus, in the strong crystal field limit, the sRDMs for crystal-field split $d$-orbitals can be obtained from the sRDMs for crystal-field split $p$ orbitals in Table.~\ref{tab:198_R}, just as at the $\Gamma$ point.

\begin{widetext}

\begin{table*}[htp]
\begin{center}
\begin{tabular}{|c|c|c|c|}
\hline
  $s$-orbital & \multicolumn{3}{|c|}{$p$-orbital}  \\
\hline
& \parbox[t]{2mm}{\multirow{3}{*}[-0.5cm]{\rotatebox[origin=c]{90}{SOC}}} & \multirow{5}{*}[0.5cm]{$P_{1/2}$} & ${\bf 0}$ \\
 & & & {$-\frac{1}{3}(\frac{m}{2}\hat{\delta k}_i +\alpha \frac{2-3|m|}{2}\lambda_{ijk}\hat{\delta k}_j\hat{\delta k}_k)~[-m]$}\\
 \cline{3-4}
 & &  & ${\bf 0}, {\bf 0}$ \\
 & & $P_{3/2}$ & {$-\frac{1}{3}(m\frac{-1+\sqrt{3}}{4}\hat{\delta k}_i +\alpha \frac{(1+\sqrt{3})(2-3|m|)}{4}\lambda_{ijk}\hat{\delta k}_j\hat{\delta k}_k)~[3m]$}\\
 ${\bf 0}$ & & & {$\frac{1}{3}(m\frac{1+\sqrt{3}}{4}\hat{\delta k}_i +\alpha \frac{(-1+\sqrt{3})(2-3|m|)}{4}\lambda_{ijk}\hat{\delta k}_j\hat{\delta k}_k)~[m]$}\\
 \cline{2-4}
 $\frac{m}{2}\hat{\delta k}_i +\alpha \frac{2-3|m|}{2}\lambda_{ijk}\hat{\delta k}_j\hat{\delta k}_k~[m]$ & \parbox[t]{2mm}{\multirow{3}{*}[-0.2cm]{\rotatebox[origin=c]{90}{Crystal field}}} & \multirow{5}{*}[0.5cm]{$p_x+p_y+p_z$} & ${\bf 0}$ \\
 & & & $\frac{m}{2}\hat{\delta k}_i +\alpha \frac{2-3|m|}{2}\lambda_{ijk}\hat{\delta k}_j\hat{\delta k}_k~[m]$\\
 \cline{3-4}
 & &  &  ${\bf 0}$, ${\bf 0}$ \\
 & & $p_x-2p_y+p_z, p_x-p_z$& $m\frac{-1+\sqrt{3}}{4}\hat{\delta k}_i +\alpha \frac{(1+\sqrt{3})(2-3|m|)}{4}\lambda_{ijk}\hat{\delta k}_j\hat{\delta k}_k~[-3m]$\\ 
 & & & $-(m\frac{1+\sqrt{3}}{4}\hat{\delta k}_i +\alpha \frac{(-1+\sqrt{3})(2-3|m|)}{4}\lambda_{ijk}\hat{\delta k}_j\hat{\delta k}_k)~[-m]$\\  
 \hline
\multicolumn{4}{|c|}{$d$-orbital}\\
\hline
& & \multicolumn{2}{|c|}{{\bf 0}, {\bf 0}}\\
 & $D_{3/2}$ & \multicolumn{2}{|c|}{$\frac{1}{5}( m\frac{-1+\sqrt{3}}{4}\hat{\delta k}_i +\alpha \frac{(1+\sqrt{3})(2-3|m|)}{4}\lambda_{ijk}\hat{\delta k}_j\hat{\delta k}_k)~[-3m]$}\\
\parbox[t]{2mm}{\multirow{3}{*}[-0.cm]{\rotatebox[origin=c]{90}{SOC}}} &  & \multicolumn{2}{|c|}{$-\frac{1}{5}(m\frac{1+\sqrt{3}}{4}\hat{\delta k}_i +\alpha \frac{(-1+\sqrt{3})(2-3|m|)}{4}\lambda_{ijk}\hat{\delta k}_j\hat{\delta k}_k)~[-m]$}\\
\cline{2-4}
& \multirow{5}{*}[0.cm]{$D_{5/2}$} & \multicolumn{2}{|c|}{{\bf 0}, {\bf 0}, {\bf 0}}\\
 & & \multicolumn{2}{|c|}{$-\frac{1}{3}(\frac{m}{2}\hat{\delta k}_i +\alpha \frac{2-3|m|}{2}\lambda_{ijk}\hat{\delta k}_j\hat{\delta k}_k)~[-m]$}\\
  &  & \multicolumn{2}{|c|}{$\frac{7}{15}( m\frac{-1+\sqrt{3}}{4}\hat{\delta k}_i +\alpha \frac{(1+\sqrt{3})(2-3|m|)}{4}\lambda_{ijk}\hat{\delta k}_j\hat{\delta k}_k)~[-3m]$}\\
   &  & \multicolumn{2}{|c|}{$-\frac{7}{15}(m\frac{1+\sqrt{3}}{4}\hat{\delta k}_i +\alpha \frac{(-1+\sqrt{3})(2-3|m|)}{4}\lambda_{ijk}\hat{\delta k}_j\hat{\delta k}_k)~[-m]$}\\
\hline
\end{tabular}
\end{center}
\caption{The sRDM for space group $P2_13 $ (198) at R-point. 
For six-fold degeneracies with states parameterized by $\alpha, m$, the $i$-th component of the sRDM $n_i(\delta\mathbf{k})$ and the winding numbers are explicitly indicated. 
For two-fold degeneracies, it turns out the sRDMs are all zero, which are indicated by ${\bf 0}$ and their winding numbers are omitted. 
When crystal field splitting dominates, the sRDMs induced from the $d$-orbitals can be determined from the corresponding rows for the $p$-orbitals, as described in the main text. An asterisk is used to denote cases where the invariant Hamiltonian has the form $\delta\mathbf{k}\cdot\mathbf{J}^*$.
}
\label{tab:198_R}
\end{table*}%

\begin{figure}[h]
  \begin{minipage}[c]{0.77\textwidth}
    \includegraphics[width=\textwidth]{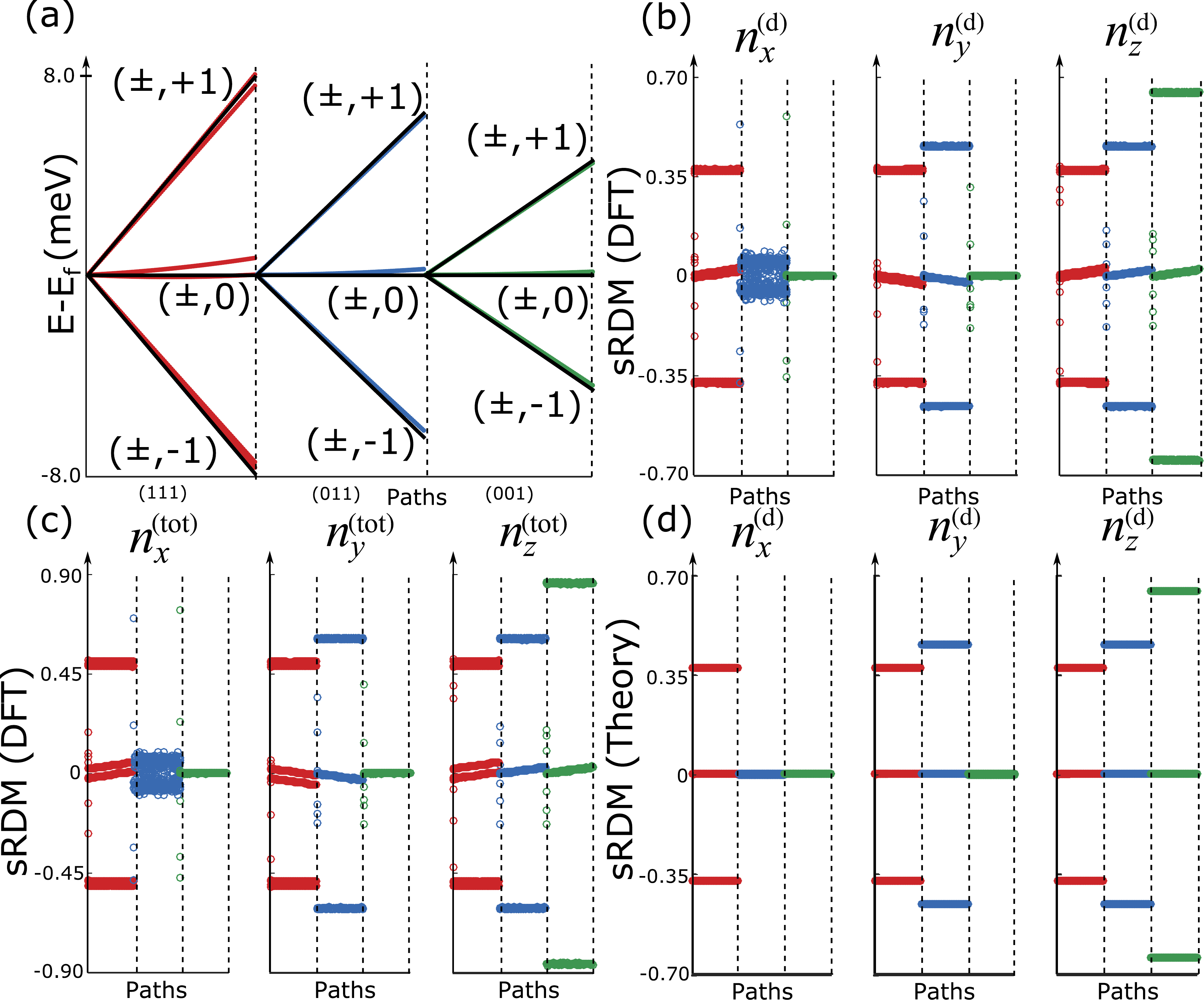}
  \end{minipage}\hfill
  \begin{minipage}[c]{0.22\textwidth}
\caption{Comparison of predicted spin-momentum locking with results from ab-initio DFT calculations. 
(a) The band structure of PtGa near the R point, focusing on the sixfold degeneracy closest to the Fermi level. Bands for $k$ vectors along (111), (011) and (001) are shown in red, blue and green respectively.
The quantum numbers $(\alpha,m)$ for the bands are shown explicitly, and the bands for distinct $\alpha$ are nearly degenerate.
The black solid lines are the fitted result from Eq.~\eqref{eq:6_fold_H}, with $hv_f/a=2.4$ meV, $\varphi = 1.4820$ and $b$ given in Eq.~\ref{eq:relation_b_main}.
(b) The DFT result for sRDM for orbitals $d_{xy}, d_{yz}, d_{zx}$, shown for different components. The colors correspond to the paths in (a).
(c) The DFT result for sRDM with all orbitals ($s$, $p$, and $d$-orbitals) included. 
(d) The $\alpha$-averaged sRDMs for the same orbitals as in (b). 
}
    \label{fig:comparison_v2}
  \end{minipage}
\end{figure}
\end{widetext}

\section{Comparision with Density Functional Theory}\label{sec:abinit}
Recently, PtGa in space group $P2_13 $ (198) has been shown experimentally to host chiral multifold fermions\cite{yao2020observation}. 
In this section, we will combine DFT calculations with our theory of sRDMs to study the spin texture of the chiral sixfold degeneracy closest to the Fermi level at the $R$ point in PtGa.
To study its electronic structure, we performed DFT calculations as implemented in the Vienna ab initio simulation package (VASP)\cite{VASP1,VASP2,VASP3,VASP4}. We use the structural parameters as reported in Ref.~\cite{Struc_PtGa}. The interaction between ion cores and valence electrons are treated by the projector augmented-wave method\cite{PAW}, the generalized gradient approximation (GGA) for the exchange-correlation potential with the Perdew-Burke-Ernkzerhof for solids parameterization\cite{PBE} and spin-orbit coupling are taken into account by the second variation method\cite{DFT-SOC}. We used a Monkhorst-Pack $\mathbf{k}$-point grid centered at $\Gamma$ of $(9\times9\times9)$ points for reciprocal space integration and 500 eV energy cutoff of the plane-wave expansion. We will in this section measure $\mathbf{k}$ vectors in reduced coordinates, so that each component of $\mathbf{k}$ lies between $0$ and $1$.

We start by plotting the calculated ab-initio band structure.
In order to fully characterize the sRDMs near the $R$-point, we select three paths, namely 
$(111)$, $(011)$ and $(001)$,
which are along the $R-\Gamma$, $R-X$ and $R-M$ directions respectively (we adopt the standard labelling of the high symmetry points in the BZ\cite{Cracknell}). 
We sample 101 points in the vicinity of the $R$-point for each path. 
Taking (111) as an example, we sample 101 data points between 
$\delta{\bf k}={\bf 0}$ (exactly at the R-point) and $\delta k_x=\delta k_y=\delta k_z=0.1$. 
For the paths (011) and (001), we keep $\delta k_x=0$ and $\delta k_x=\delta k_y=0$ respectively. 
We show the band structure in Fig.~\ref{fig:comparison_v2}(a); we use red for the bands along $(111)$, blue for bands along $(011)$, and green for bands along $(001)$. 
We observe that to linear order the bands appear to be doubly degenerate and linearly dispersing in all directions. 
Quadratic corrections to the dispersion can be observed for the middle two bands along all three paths. 
We also attribute the small splitting of the bands along the $(111)$ direction to the quadratic corrections, since the splitting is of the same order as the deviation from linear dispersion along the $(011)$ and $(001)$ directions. Furthermore, as shown in Ref.~\cite{flicker2018chiral}, if the bands of a sixfold fermion are degenerate to linear order along $(111)$, then they are degenerate to linear order for every $\delta\mathbf{k}$. Although this double degeneracy is not symmetry protected (and hence nonuniversal), it is similar to the dispersion in other B20 compounds such as RhSi, and arises due to the dominance of short-range hoppings\cite{chang2017unconventional}. 
The bands in Fig.~\ref{fig:comparison_v2}(a) are similar to what we have analyzed in Sec.~\ref{sec:198_R}, where we specialized to the exactly solvable case of Eq.~\eqref{eq:6_fold_H} with $\varphi=\pi/2, b=0$. 
However, we cannot deduce that $b=0,\varphi=\pi/2$ from the dispersion alone. 
In order to make quantitative comparison with the DFT result, we show in Appendix~\ref{sec:alpha-averaged} that if
\begin{align}
\label{eq:relation_b_main}
    b = \sqrt{\cos^2\varphi-3\sin^2\varphi},
\end{align}
then the eigenvalues of $H^{(6)}(\delta{\bf k}, b, \varphi, v_f)$ are doubly degenerate and given by
\begin{align}
\label{eq:double_spectrum_main}
    \{0, \pm 2\hbar v_f\cos\varphi|{\bf \delta k}|\}. 
\end{align}
Eq.~\eqref{eq:double_spectrum_main} can be viewed as a first order approximation for the DFT band structures shown in Fig.~\ref{fig:comparison_v2}(a), with two to-be-determined parameters $v_f$ and $\varphi$.
Since the sRDMs depend on $\varphi$ and not $v_f$, we can first determine the value of $\varphi$ by computing the sRDM from DFT at a single $\mathbf{k}$ point. 
We can then fit $v_f$ to the band dispersion. 
Following this, we can compare our theoretically determined sRDM to the spin texture computed from DFT at other $\mathbf{k}$ points.

The sRDMs can be calculated from the ab-initio data for each sampled data point on a given band $n$. 
To do so, we compute the projection of the DFT wave functions onto the spherical harmonics $|Y_{lm}\rangle$ and collect the results into an ``orbital weight table''\cite{schuler2018charge}. 
The orbital weight table consists of four sectors, $j=0,1,2,3$, consisting of the average of the identity operator $j=0$, spin $s_x$ ($j=1$), spin $s_y$ ($j=2)$ and spin $s_z$ ($j=3$) of the projected wave functions. 
We can organize each sector into a matrix $M_j$ with matrix element $M_j^{\beta,lm}$. 
Here $\beta=1,...,8$ labels the eight ions in the unit cell, and $lm$, which should be treated as a single index, labels the orbitals. 
In App.~\ref{sec:Calculation of spin reduced density matrix from the orbital weight table}, we show that the spin-resolved density matrix for a given orbital $lm$ and ion $\beta$, is given by 
\begin{eqnarray}\begin{aligned}
\label{eq:def_spinRDM_main_before}
\rho_{\delta\bf k}^{\beta,lm}=\frac{1}{2}\sum_{j=0}^3M^{\beta,lm}_j\sigma^j,
\end{aligned}\end{eqnarray}
where $\sigma^j$ are the four Pauli matrices (we define $\sigma^0$ to be the $2\times2$ identity matrix). 
The \emph{reduced} density matrix is obtained by tracing out the ionic and orbital degrees of freedom
\begin{eqnarray}\begin{aligned}
\label{eq:def_spinRDM_main}
\rho_{\delta\bf k}^{\text{RDM}} \equiv \sum_{\beta,lm}\rho_{\delta\bf k}^{\beta,lm},
\end{aligned}\end{eqnarray}
such that the sRDM can be directly extracted from the DFT data as
\begin{align}
\label{eq:def_spinRDM_main_2}
    n_i({\delta\bf k}) = \text{tr}\left(\rho_{\delta\bf k}^{\text{RDM}}\sigma_i\right).
\end{align}
PtGa was found to have the strongest SOC among all the chiral multifold fermionic materials investigated thus far\cite{yao2020observation}.
From the orbital weight table, we compute that the dominant contribution ($\sim74$\%) to bands near the sixfold degeneracy comes from the orbitals $d_{xy}, d_{yz}$ and $d_{zx}$.  
The sRDMs from these orbitals, calculated with Eq.~\eqref{eq:def_spinRDM_main}-\eqref{eq:def_spinRDM_main_2} are shown in Fig.~\ref{fig:comparison_v2}(b).
Here, the $x,y$ and $z$ components of the sRDMs are shown in separate panels, with the red, blue and green circles correspond to paths $(111)$, $(011)$ and $(001)$ respectively. 
As a comparison, the sRDMs from all the orbitals, including the small contributions from $s$, $p$ and $d$-orbitals, are shown in Fig.~\ref{fig:comparison_v2}(b).
Upon comparison, it is clear that the sRDMs from the $d_{xy},d_{yz},$ and $d_{zx}$ orbitals constitute a good approximation to the total sRDM, and so we will focus on them for the remainder of the discussion. 
The sRDMs from other orbitals will be discussed in details in App.~\ref{sec:Calculation of spin reduced density matrix from the orbital weight table}. 

We can theoretically calculate the sRDMs induced from the above $d$-orbitals using the approach developed in Sec.~\ref{sec:198_R}. 
From the DFT calculated orbital weights, we observe that $d_{xy}, d_{yz}$ and $d_{zx}$ have equally distributed weights, which hints that 
crystal field splitting dominates over on-site SOC, and that the dominant contribution of sRDM comes from the $d_{xy} + d_{yz} + d_{zx}$ orbitals of Pt at the $4a$ Wyckoff position. 
Since this orbital behaves like the $p_x+p_y+p_z$ orbital (see Table~\ref{tab:207_p_d_symmetry}), we can use the basis of Eq.~\eqref{eq:rpoint6foldspace} with the Wyckoff positions $|{\bf q}_i\rangle$ replaced by the ``dressed'' Wyckoff positions $|\bar{\bf q}_i\rangle$ defined in Eq.~\eqref{eq:dress_wyckoff}. 
The eigenstates of the sixfold Hamiltonian $H_R^{(6)}(\delta{\bf k}, b, \varphi, v_f)$, which can be labeled as $|\alpha=\pm, \delta{\bf k}m\rangle$, serve as coefficients to determine a linear combination of these basis states, from which we can extract the sRDM ${\bf n}^{\alpha,m}$.
Provided Eq.~\eqref{eq:relation_b_main} holds, the states $|\alpha=\pm, \delta{\bf k}m\rangle$ are doubly degenerate with $\alpha$-independent eigenenergy.
As a result, we expect that the ab-initio calculated sRDMs are averaged sRDM over $\alpha$, as discussed in Sec.~\ref{sec:198_R_s}. In the $\delta\mathbf{k}\cdot\mathbf{J}$ limit with $b=0,\phi=\pi/2$, the sRDM is given by the ``Crystal field'' $p_x+p_y+p_z$ row of Table~\ref{tab:198_R}, which coincides with Eq.~\eqref{eq:spinRDM_198_R_6fold} for s orbitals. To go beyond the $\delta\mathbf{k}\cdot\mathbf{J}$ result quantitatively, we must fit $v_f$ and $\phi$ to the DFT using Eq.~\eqref{eq:relation_b_main}.

With this understanding, we scan through $\varphi\in[-\pi, \pi)$, diagonalize $H_R^{(6)}(\delta{\bf k}, b, \varphi, v_f)$ (recall that $v_f$ would not affect sRDM), followed by calculating the $\alpha$-averaged sRDM for the $d_{xy} + d_{yz} + d_{zx}$ orbital. 
After comparing to $n_z^{(d)}$ along the $(001)$ direction in Fig.~\ref{fig:comparison_v2}(b), we determine that  $\varphi=1.4820$ (see Appendix~\ref{sec:alpha-averaged} for more details). 
We can then use this value of $\varphi$ to determine the sRDM along the other paths, shown in Fig.~\ref{fig:comparison_v2}(d) . 
With the value of $\varphi$ determined, we can determine $v_f$ by fitting the band structures in Eq.~\eqref{eq:double_spectrum_main} to that from the DFT result. 
We find that $hv_f/a=2.4$ meV gives the best fit, where $a=4.973\text{\AA}$  is the lattice constant for PtGa. 
The bands computed with this value of $v_f$ are shown as black solid lines in  Fig.~\ref{fig:comparison_v2}(a). 

From Fig.~\ref{fig:comparison_v2}, we first notice that the sRDMs are approximately constant along each path
for $|\delta {\bf k}|a<8 \text{meV}/\hbar v_f$,
which is a priori not obvious from Eq.~\eqref{eq:def_spinRDM_main_before}-\eqref{eq:def_spinRDM_main_2}.
However, from Eq.~\eqref{eq:6_fold_H}, we can always factor out the momentum for a given path such that the eigenstates and the sRDM are independent of the magnitude of the momentum. 
Thus, this observation serves as the first confirmation for our theory. 
For the (111) path, which is labeled in red, the sRDM from DFT and theoretical prediction are both isotropic, and show excellent quantitative agreement. 
This suggests that the second order correction to the band structure, which is visible from Fig.~\ref{fig:comparison_v2}(a), has negligible effect on the sRDM. 
For the path (011) labelled in blue, the x-component of the sRDM calculated from DFT is noisy due to the degeneracy of the bands; averaging over this, we have excellent quantitative agreement. 
The consistency between the DFT and theory result is also manifested for the path (001) labeled in green (although this was guaranteed by the fitting procedure). 
For both the ab-initio and the theoretically calculated sRDM, we confirm that the sRDMs for the $d_{xy}+d_{yz}+d_{zx}$ orbital are radial. 
This may seem contradictory to our results in Sec.~\ref{sec:198_R}, particularly those shown in  Table~\ref{tab:198_R}. 
However, the non-radial texture in the sRDMs arises because we have manually separated the two degenerate bands into different $\alpha$-sectors.
As discussed in Sec.~\ref{sec:198_R}, even though $b\neq0$, since the bands are nearly doubly degenerate, the more physical quantity for degenerate bands is the $\alpha$-averaged sRDMs; the non-radial $\alpha$-dependent contributions cancel in the average. %
We note that the $\alpha$-averaged sRDMs at the R-point for SG 198 are generally proportional to the magnetic quantum number $m$, as evident from Table~\ref{tab:198_R}.

In summary, with a focus on the sixfold degeneracy near the $R$-point, we show that our theory provides both quantitatively accurate predictions, as well as valuable insights to some features for the sRDMs of PtGa.
We stress that it is important to distinguish which mechanism, either on-site SOC or crystal field, is dominant and responsible for the splitting of the band representations. 
The sRDM is determined not only by the band representations, but also the underlying basis that spans the subspace of the degeneracy. 
In fact, considering that SOC is a relativistic effect, we expect that crystal field splitting should be the leading energy scale for most real material, which is also confirmed in our comparison above. 
Additionally, we have seen that the energy bands alone are not enough to uniquely specify the $\mathbf{k}\cdot\mathbf{p}$ Hamiltonian for a sixfold degenerate fermion. 
Our fitting shows that we need to additionally consider the spin of the eigenstates to remove the redundancy Eq.~\eqref{eq:relation_b_main}.

\section{A Practial Guide to the Method}\label{sec:inpractice}

In Sec.~\ref{sec:abinit} we have seen through a specific example how our method can be combined with DFT calculations to yield predictions for the spin texture near the sixfold degeneracy in PtGa. In this section, we will give a general guide on how to apply the results of this work to other materials and $\mathbf{k}$ points in space group $P2_13 $ (198). As we saw in PtGa, the first step is to perform an orbital-resolved ab initio calculation for bands near the $\mathbf{k}$ point of interest (either the $\Gamma$ or $R$ point for this work). From the orbital projection at the degeneracy point, one can read off the orbitals that compose the band degeneracies closest to the Fermi level. Since we expect that for most materials that crystal field splitting dominates over on-site SOC, we can focus on the linear combinations of crystal-field split orbitals identified in Tables~\ref{tab:198_Gamma} and \ref{tab:198_R}. Once the dominant set of crystal-field split orbitals near the Fermi level is known, the tables give heuristics for what to expect for the spin texture in the $\delta\mathbf{k}\cdot\mathbf{J}$ limit, in the absence of any type-II entanglement. To go further and make quantitative predictions for the spin texture, we can fit the bands to the low-energy $\mathbf{k}\cdot\mathbf{p}$ invariant Hamiltonian, just as we did in Sec.~\ref{sec:abinit} and Appendix~\ref{sec:dftdetails}. We can then numerically construct the pullback map, and hence the sRDM. 

Although we focused on the $\Gamma$ and $R$ point of space group $P2_13 $ (198) here, our method can be extended to any space group, and to any high-symmetry point. To do so, we need to compute the invariant $\mathbf{k}\cdot\mathbf{p}$ Hamiltonian for each allowed degeneracy at the high symmetry point. Additionally, we need to compute the elementary band representations carried by each set of (crystal-field or spin-orbit split) orbitals. Finally, for each elementary band representation we need to compute the expression for the basis vectors of each degeneracy in momentum space, analogous to our Eqs.~\eqref{eq:rpoint2foldspace} and \eqref{eq:rpoint6foldspace}. This allows for the computation of the pullback maps---and hence the sRDMs---for any space group.

\section{Discussion and Conclusion}
\label{sec:Discussion and Conclusion}

In this work we have shown how the tools of band representations and topological quantum chemistry yield insight into the origin of spin-momentum locking in multifold fermions. 
We have introduced a set of analytical tools that let us compute the bulk spin texture for nodal points in any space group. 
As a test case, we analyzed two- and fourfold degeneraces at the $\Gamma$ and $R$ point of the chiral octahedral space group $P432$ (207). 
We highlighted the importance of local (position space) basis functions in determining the spin texture near degeneracies, showing that while twofold and fourfold degeneracies from $s$ and $p$ orbitals always had radial spin textures with winding number $\pm 1$ over the Fermi surface, $d$ orbitals could yield fourfold degeneracies with spin winding number $\pm 1$ and $\pm 5$. 
We additionally applied our theory to the experimentally interesting case of B20 compounds in space group $P2_13$ (198) like AlPt, PdGa, and PtGa. 
Along the way, we have also quantified the fragility of spin momentum locking both to on-site spin-orbital entanglement and to the mixing of additional basis states. 
As an example, we have shown that the spin-orbit coupled twofold degeneracy at the $R$ point of $P2_13$ can have vanishing spin-momentum locking. 
We also derived the allowed spin texture for the sixfold degeneracy at the $R$ point, where the $\mathbf{n}$ vector winds either $1$ or $0$ times around the Fermi surface despite being non-radial. Additionally, for $p$ and $d$ orbitals we found that the sixfold degeneracy could have spin texture with winding $\pm 3$. 
We note that our results also extend to threefold degeneracies at the $P$ point in space group $I2_13$ (199), which have a $\mathbf{k}\cdot\mathbf{p}$ Hamiltonian given by $H^{(3)}$ in Eq.~\eqref{eq:6_fold_H} (the analogous threefold degeneracy in space group $I4_123$ (214) could be analyzed similarly, but one must account for the enlarged site-symmetry group at the maximal Wyckoff positions).
Beyond just theoretical analysis, our work can be combined with ab initio calculations to extract the spin-momentum locking of Bloch states near multifold nodal degeneracies, by combining basis orbital information with fitting to the $\mathbf{k}\cdot\mathbf{p}$ dispersion. 
In contrast to direct magnetization calculation in DFT, our method allows for systematic approximation schemes by excluding type-II entangled orbitals. 
Additionally, we have shown that our theory allows for a quantitatively accurate prediction of the spin texture along any $\mathbf{k}$ direction using ab-initio data from only a few momenta.

A key point we emphasize is that spin-momentum locking is not a quantized concept, even in cases where simple models predict a winding of the spin texture around the Fermi surface. 
Indeed, the vector $\mathbf{n}(\mathbf{k})$ which determines the sRDM is \emph{not} a vector of constant length. 
We have shown that both on-site (type-I) and multisite (type-II) SOC-induced entanglement can affect the magnitude and the direction of $\mathbf{n}(\mathbf{k})$; whenever a perturbation pushes the length $|\mathbf{n}(\mathbf{k})|$ to zero, all spin-momentum locking is destroyed and even spin-textures that appear to have nontrivial winding can be unwound. 

Our work opens the door to several interesting experimental and theoretical questions. 
First, our predictions suggest that bulk spin-resolved ARPES in chiral topological semimetals could reveal interesting spin textures related to the winding number of topological degeneracies. 
However, our work shows that care must be used in interpreting the results due to both the nonquantization of spin-momentum locking, as well as the dense sampling in momentum space necessary to infer any winding of the spin texture. 
Additionally, our work suggests that materials with chiral multifold fermions may exhibit chirality-induced spin selectivity and spin transport beyond what has recently been predicted for simple Kramers-Weyl fermions\cite{roy2022long,calavalle2022gate,shiota2021chirality}. 
Next, our method should also be applicable in a suitably generalized form to the surface Fermi arc states in topological semimetals. 
A pressing question is to what degree the Fermi arc states inherit the spin-mometnum locking from the bulk, and furthermore how robust this is to surface effects. 
This will be the subject of a forthcoming work.
Finally, we showed that fourfold degeneracies can host spin textures with winding number $\pm 1, \pm 5,$ and even $\pm 7$. 
This raises the intriguing question of what happens to spin-dependent interactions such as triplet pairing or magnetic interaction when projected into a Fermi surface with large spin texture. 
Our work suggests that superconducting instabilities in chiral multifold fermions may exhibit a rich structure due to this spin-momentum locking.

\section{Acknowledgements}       
M.G.V. thanks support to Programa Red Guipuzcoana de Ciencia Tecnología e Innovación 2021 No. 2021-CIEN-000070-01 Gipuzkoa Next and the Deutsche Forschungsgemeinschaft (DFG, German Research Foundation) GA 3314/1-1 – FOR
5249 (QUAST). M.G.V. and I.R. acknowledge the Spanish Ministerio de Ciencia e Innovacion (grant PID2019-109905GB-C21). B.~B. acknowledges the support of the Alfred P. Sloan foundation, and the National Science Foundation under grant DMR-1945058.

\appendix

\section{Some Useful Character Tables}
\label{sec:Some Useful Character Tables}
In this section, we provide the character tables for groups used in this work. 
We use the notation of Refs.~\cite{Bilbao1,Bilbao2,Bilbao3,elcoro2017double} for the representation labels, and the properties of the basis functions can be found in, e.~g., Ref.~\cite{PointGroupTables}. 
First, we consider the chiral octahedral group $432$ ($O$), which is the point group of space group $207$. 
It has eight irreducible representations, five spinless and three spinful. 
Each of the representations is also time-reversal invariant. 
They are defined in Table~\ref{tab:432chars}. 
We note that the trivial $A$ representation is carried by spinless $s$ orbitals. 
The $T_1$ representation is carried by spinless $p$ orbitals. 
The triplet of spinless $(d_{xz},d_{yz},d_{xy})$ orbitals carry the $T_2$ representation. 
The pair of $(d_{x^2-y^2},d_{z^2})$ carry the $E$ representation. 
For spinful orbitals, we note that $J=1/2$ states transform in the $\bar{E}_1$ representation, while $J=3/2$ states transform in the $\bar{F}$ representation. 
Finally, the $|Jm_j\rangle=\ket{5/2\pm 5/2}$ states transform in the $\bar{E}_2$ representation, while the remaining four $J=5/2$ states transform in the $\bar{F}$ representation.
\begin{table}[t]
\begin{tabular}{c|c|c|c|c|c|c}
$\rho$ & $E$ & $C_{2z}$ & $C_{31}$ & $C_{2a}$ & $C_{4z}$ & $\bar{E}$ \\
\hline
$A_1$ & 1 & 1 & 1 & 1 & 1 & 1 \\
$A_2$ & 1 & 1 & 1 & -1 & -1 & 1 \\
$E$ & 2 & 2 & -1 & 0 & 0 & 2 \\
$T_1$ & 3 & -1 & 0 & -1 & 1 & 3 \\
$T_2$ & 3 & -1 & 0 & 1 & -1 & 3 \\
$\bar{E}_1$ & 2 & 0 & 1 & 0 & $\sqrt{2}$ & -2 \\
$\bar{E}_2$ & 2 & 0 & 1 & 0 & $-\sqrt{2}$ & -2 \\
$\bar{F}$ & 4 & 0 & -1 & 0 & 0 & -4
\end{tabular}
\caption{Character table for the point group $432$. 
The first column gives the representation label. 
Each subsequent column gives the character for elements in the conjugacy class with representative elemnt given in the first column.}\label{tab:432chars}
\end{table}

Next, in Table~\ref{tab:c3chars}, we give the character table for $C_3$, the site-symmetry group of the $4a$ position in space group $198$. 
Unlike in $432$, here not all of the irreps are time-reversal invariant. 
Time-reversal pairs the ${}^1E$ and ${}^2E$ irrep into a ``physically irreducible'' representation we denote as ${}^1E{}^2E$. 
Similarly, for spinful orbitals the physically irreducible representations are $\bar{E}\bar{E}$ and ${}^1\bar{E}{}^2\bar{E}$. 
For concreteness, let us focus on the case where the $C_3$ axis is in the $111$ axis, as occurs in space group $198$. 
In this case, $A_1$ representation is carried by spinless $s$ orbitals, and by the spinless $p_x+p_y+p_z$ orbital. 
The ($p_x -2p_y+p_z$, $p_x-p_z$) orbitals carry the physically irreducible ${}^1E{}^2E$ representation. 
Turning to spinful orbitals, $S_{1/2}$ and $P_{1/2}$ orbitals each carry the ${}^1\bar{E}{}^2\bar{E}$ physically irreducible representation. 
In fact, choosing the angular momentum quantization axis along $111$, we have that each pair of $m_J=\pm (4n+1)/2$ states carry this representation, for $n=0,1,\dots$. 
Alternatively, we have that $m_J=\pm (4n-1)/2$ states carry the $\bar{E}\bar{E}$ representation, for $n=1,2,\dots$.

\begin{table}[t]
\begin{tabular}{c|c|c|c|c}
$\rho$ & $E$ & $C_3$ & $C_3^{-1}$ & $\bar{E}$ \\
\hline
$A_1$ & 1 & 1 & 1 & 1\\
${}^2E$ & 1 & $e^{-2\pi i/3}$ & $e^{2\pi i/3}$ & 1 \\
${}^1E$ & 1 &  $e^{2\pi i/3}$ & $e^{-2\pi i/3}$ & 1 \\
$\bar{E}$ & 1 & -1 & -1 & -1 \\
${}^1\bar{E}$ & 1 & $e^{-\pi i/3}$ & $e^{\pi i/3}$ & -1 \\
${}^2\bar{E}$ & 1 & $e^{\pi i/3}$ & $e^{-\pi i/3}$ & -1 \\
\end{tabular}
\caption{
Character table for the point group $C_3$. 
The first column gives the representation label. 
Each subsequent column gives the character for elements in the conjugacy class with representative elemnt given in the first column.}\label{tab:c3chars}
\end{table}
\section{The band representations for the $(p_x-2p_y+p_z, p_x-p_z)$ orbitals in the strong crystal field limit}
\label{sec:bandrep_p_crystal}
In this section, we shall explicitly construct the band representation matrices in the subspace spanned by the $(p_x-2p_y+p_z, p_x-p_z)$-orbitals in the limit of strong crystal field splitting for space group $P2_13$ (198). 
We shall first introduce the ``dressed'' Wyckff positions in this subspace, followed by showing how the band represetnation matrices split at the $\Gamma$ and $R$ points. 
In Sec.~\ref{sec:198_Gamma}, we have explicitly introduced the ``dressed'' Wyckoff positions for the $p_x+p_y+p_z$ subspace (See Eq.~\eqref{eq:dress_wyckoff}); 
now we introduce the analogous basis for the $(p_x-2p_y+p_z, p_x-p_z)$ subspace as
\begin{eqnarray}
\begin{aligned}
\label{eq:dress_wyckoff_2}
 |\bar{\bf q}_5\rangle &\equiv |{\bf q}_1\rangle\otimes\frac{|p_x\rangle-|p_z\rangle}{\sqrt{2}}, \\
 |\bar{\bf q}_6\rangle &\equiv |{\bf q}_2\rangle\otimes\frac{|p_x\rangle+|p_z\rangle}{\sqrt{2}}
 =\rho_{{\bf k}, W\otimes O}(C_{2x})|\bar{\bf q}_5\rangle, \\
 |\bar{\bf q}_7\rangle &\equiv |{\bf q}_3\rangle\otimes\frac{-|p_x\rangle-|p_z\rangle}{\sqrt{2}}
 \\&
 =\rho_{{\bf k}, W}(C_{2y})\otimes\rho_{{\bf k}, O}(C_{2z})|\bar{\bf q}_5\rangle, \\
 |\bar{\bf q}_8\rangle &\equiv |{\bf q}_4\rangle\otimes\frac{-|p_x\rangle+|p_z\rangle}{\sqrt{2}}
 \\&
 =\rho_{{\bf k}, W}(C_{2z})\otimes\rho_{{\bf k}, O}(C_{2y})|\bar{\bf q}_5\rangle\\
 |\bar{\bf q}_9\rangle &\equiv |{\bf q}_1\rangle\otimes\frac{|p_x\rangle-2|p_y\rangle+|p_z\rangle}{\sqrt{6}}, \\
 |\bar{\bf q}_{10}\rangle &\equiv |{\bf q}_2\rangle\otimes\frac{|p_x\rangle+2|p_y\rangle-|p_z\rangle}{\sqrt{6}}
 =\rho_{{\bf k}, W\otimes O}(C_{2x})|\bar{\bf q}_9\rangle, \\
 |\bar{\bf q}_{11}\rangle &\equiv |{\bf q}_3\rangle\otimes\frac{-|p_x\rangle+2|p_y\rangle+|p_z\rangle}{\sqrt{6}}
 \\&
 =\rho_{{\bf k}, W}(C_{2y})\otimes\rho_{{\bf k}, O}(C_{2z})|\bar{\bf q}_9\rangle, \\
 |\bar{\bf q}_{12}\rangle &\equiv |{\bf q}_4\rangle\otimes\frac{-|p_x\rangle-2|p_y\rangle-|p_z\rangle}{\sqrt{6}}
 \\&
 =\rho_{{\bf k}, W}(C_{2z})\otimes\rho_{{\bf k}, O}(C_{2y})|\bar{\bf q}_9\rangle.
\end{aligned}
\end{eqnarray}
We have constructed the basis by acting with twofold rotations on the two orbitals $\ket{\bar{\mathbf{q}}_5}$ and $\ket{\bar{\mathbf{q}}_9}$ centered at $\mathbf{q}_1$; unlike in Eq.~\eqref{eq:dress_wyckoff} we now have a two-dimensional subspace on each site. 
We can project the spinless band representation matrices into this basis, and the generators read
\begin{eqnarray}\begin{aligned}
\label{eq:bandrep_8}
    &\rho_{\bf k}^{(8)}(C_{31}^+)=\begin{pmatrix}
-\frac{1}{2} & -\frac{\sqrt{3}}{2} \\
\frac{\sqrt{3}}{2} & -\frac{1}{2}
\end{pmatrix}\otimes\left(\begin{array}{cccc}
    1 & 0 & 0 & 0 \\
    0 & 0 & 0 & 1 \\
    0 & 1 & 0 & 0 \\
    0 & 0 & 1 & 0
    \end{array}\right),\\
    &\rho_{\bf k}^{(8)}(\{C_{2x}|\half\half 0\})=\mathbb{I}_{2}\otimes
    \left(\begin{array}{cccc}
    0 & \pm1 & 0 & 0 \\
    1 & 0 & 0 & 0 \\
    0 & 0 & 0 & \pm1 \\
    0 & 0 & 1 & 0
    \end{array}\right),\\
    &\rho_{\bf k}^{(8)}(TR)=i\mathbb{I}_{2}\otimes\mathbb{I}_4\mathcal{K},
\end{aligned}\end{eqnarray}
for $\mathbf{k}\in\{\Gamma,R\}$, and the signs in $\rho_{\bf k}^{(8)}(\{C_{2x}|\half\half 0\})$ is $+1$ if $\mathbf{k}=\Gamma$ and $-1$ if $\mathbf{k}=R$. 
We notice the second terms in the tensor products are nothing but $\rho_{{\bf k},W}(g)$ in Eq.~\eqref{eq:bandrep_Wyckoff}. 
Hence at the $\Gamma$-point, they can be block-diagonalized as in Eq.~\eqref{eq:split_wyckoff_gamma} into a one dimensional $\ket{\bar s}$ and a three dimensional $\ket{\bar p}$representation. 
Upon tensoring with the two-dimensional Hilbert space of spin at each site, and using the addition of angular momentum we find the Hilbert space dimensions split as
\begin{align}
    {\bf 2}\otimes({\bf 1}\oplus{\bf 3})\otimes{\bf 2} = {\bf 4}\oplus  {\bf2}\otimes({\bf 2}\oplus{\bf 4}) = {\bf 4}\oplus {\bf 4}\oplus ({\bf 2}\oplus{\bf 2}\oplus{\bf 4}),
\end{align}
where we have used effective SOC twice to decouple the irreps. 
This explains the dimensions of the degeneracies for the rows labelled by ``$p_x-2p_y+p_z, p_x-p_z$'' in Table.~\ref{tab:198_Gamma}. Applying this decomposition directly to the matrices in Eq.~\eqref{eq:bandrep_8} gives us the pullback maps needed to determine the sRDMs.

For the $R$-point, similarly, upon coupling the ``Wyckoff'' parts in Eq.~\eqref{eq:bandrep_8} to the spin, they split into a two-dimensional and another six-dimensional representations, as in Eq.~\eqref{eq:bandrep_198_R}. 
We can keep track of the dimensions as 
\begin{align}
    {\bf 2}\otimes({\bf 4}\otimes{\bf 2}) = {\bf 2}\otimes({\bf 2}+{\bf 6}) = {\bf 4}\oplus {\bf 12} = ({\bf 2}\oplus {\bf 2}) \oplus({\bf 6}\oplus {\bf 6} )
\end{align}
where in the last line, we solved the constraint equation Eq.~\eqref{eq:constraint_eqn_more_general} in the subspaces denoted by ${\bf 4}$ and ${\bf 12}$ separately. 
Thus we find two twofold and sixfold degeneracies respectively as shown in Table.~\ref{tab:198_R}. 

\section{Details for comparing theory and DFT results for sRDMs}\label{sec:dftdetails}
In this appendix, we illustrate several techniques used in comparing our theoretically calculated sRDM to the results of ab-initio calculations. 

\subsection{Calculation of spin reduced density matrix from the orbital weight table}
\label{sec:Calculation of spin reduced density matrix from the orbital weight table}

First, we demonstrate how to obtain the spin reduced density matrix for a given band $n$ from the orbital weight table computed from DFT. 
Suppose for a given band n, the wavefunction near the R-point can be represented as follows
\begin{eqnarray}\begin{aligned}\label{eq:def_Psi}
|\Psi_{n}\rangle = \begin{bmatrix}
\chi_{nk}^\uparrow(1) & \chi_{nk}^\downarrow(1) &... &\chi_{nk}^\uparrow(8) & \chi_{nk}^\downarrow(8) 
\end{bmatrix}^T,
\end{aligned}\end{eqnarray}
where $\chi_{nk}^\mu(\beta)$ is the component of the wavefunction with spin $\mu$ at the $\beta$-th ion. 
For the case of PtGa, $\beta=1,...,8$ since there are eight ions per unit cell, but our analysis is completely general. We assume that the wave functions are normalized, such that 
$\langle\Psi_n|\Psi_m\rangle = \delta_{mn}$.
The orbital weight table consists of the projections
\begin{eqnarray}\begin{aligned}
\label{eq:matrix_elemnt_1}
\langle Y_{lm}|\left(\sum_{\mu,\nu=1}^2|\chi^\nu_{n{\bf k}}(\beta)\rangle\sigma^j_{\mu\nu}\langle\chi^\mu_{n{\bf k}}(\beta)|\right)|Y_{lm}\rangle ,
\end{aligned}\end{eqnarray}
onto spherical harmonics for $j=0,1,2,3$, given in four rows. 
Here $lm$ (which should be treated as one index such as $s$, $p_x$, $d_{x^2-y^2}$, etc) label the orbitals, and $\beta$ labels the ions. 
$\mu,\nu=1,2$ correspond to spin up and spin down respectively.
We note that $\langle Y_{lm}|\chi^\mu_{nk}(\beta)\rangle$ is the projection of the wavefunction onto a spherical harmonic and spin $\mu$ at the $\beta$-th ion. 
Thus, we introduce the following notation 
\begin{eqnarray}\begin{aligned}
\langle Y_{lm}|\chi^\mu_{n{\bf k}}(\beta)\rangle = z_\mu^{\beta,lm} ,
\end{aligned}\end{eqnarray}
for a fixed band $n$ (for brevity we omit the $n$ and ${\bf k}$ labels). We can then simplify Eq.~\eqref{eq:matrix_elemnt_1} as
\begin{eqnarray}\begin{aligned}
\label{eq:matrix_elemnt_3}
\sum_{\mu,\nu=1,2}(z_\mu^{\beta,lm})^*\sigma_{\mu\nu}^j (z_\nu^{\beta,lm}), &\quad j =  0,1 ,2 ,3  ,
\end{aligned}\end{eqnarray}
which reads explicitly
\begin{eqnarray}\begin{aligned}
\label{eq:matrix_elemnt_2}
j = 0 : & \left( |z_1^{\beta,lm}|^2 + |z_2^{\beta,lm}|^2\right); \\
j = 1 : &\left(z_1^{\beta,lm*}z_2^{\beta,lm}+h.c.\right); \\
j = 2 : &\left(-iz_1^{\beta,lm*}z_2^{\beta,lm}+h.c.\right); \\
j = 3 : & \left( |z_1^{\beta,lm}|^2 - |z_2^{\beta,lm}|^2 \right).
\end{aligned}\end{eqnarray}
For each sector $j$, we see that we can view the given orbital weights as a matrix $M_j$, with matrix element $M_j^{\beta,lm}$ for ion $\beta$ and orbital $lm$. 
The matrix elements are given in Eq.~\eqref{eq:matrix_elemnt_3} as
\begin{eqnarray}\begin{aligned}
M_j^{\beta,lm} = \sum_{\mu,\nu=1,2}(z_\mu^{\beta,lm})^*\sigma_{\mu\nu}^j (z_\nu^{\beta,lm}),
\end{aligned}\end{eqnarray}
which satisfy
\begin{eqnarray}\begin{aligned}
\label{eq:matrix_elemnt_constraint}
M_{j=0}^{\beta,lm} = \sqrt{(M_{j=1}^{\beta,lm})^2+(M_{j=2}^{\beta,lm})^2+(M_{j=3}^{\beta,lm})^2},
\end{aligned}\end{eqnarray}
for each entry $(\beta,lm)$. 
Other than this property, we also note from the first line in Eq.~\eqref{eq:matrix_elemnt_2} that all the entries in $M_{j=0}$ are positive. 
Furthermore, the spherical harmonics satisfy 
\begin{align*}
\langle Y_{l'm'}|Y_{lm}\rangle =\delta_{ll'}\delta_{mm'},
\end{align*}
and are taken by the ab-initio code to be orthogonal for atoms at different sites\cite{schuler2018charge}. The orbital weight table is normalized such that
\begin{align}
\label{eq:ortho}
\sum_\beta\sum_{lm}\sum_{\mu=1}^2 \big|\langle Y_{lm}|\chi^\mu_{n{\bf k}}(\beta)\rangle\big|^2 = \sum_\beta\sum_{lm}\sum_{\mu=1}^2 |z_\mu^{\beta,lm}|^2  = 1,
\end{align}
where we have taken into account two spin species. 
In other words, the sum of all entries with $j=0$ is unity.

The spin-resolved  density matrix for a given orbital $lm$ and ion $\beta$ is denoted as $\rho^{\beta,lm}$, and it has $\mu\nu$-components defined as (note the switch of indices)
\begin{eqnarray}\begin{aligned}
\label{eq:density_matrix}
\rho^{\beta,lm}_{\mu\nu} &\equiv (z^{\beta,lm}_{\nu})^*(z^{\beta,lm}_{\mu}) \\
&= \begin{bmatrix}
|z_1^{\beta,lm}|^2 & (z^{\beta,lm}_2)^*(z^{\beta,lm}_1)  \\
(z^{\beta,lm}_2)(z^{\beta,lm}_1)^* & |z^{\beta,lm}_2|^2
\end{bmatrix},
\end{aligned}\end{eqnarray}
which has trace one by virtue of Eq.~\eqref{eq:ortho}.
We observe that Eq.~\eqref{eq:matrix_elemnt_3} can be rewritten as
\begin{eqnarray}\begin{aligned}\label{eq:dm_from_weights}
\rho^{\beta,lm}_{\mu\nu}=\frac{1}{2}\sum_{j=0}^3M^{\beta,lm}_j\sigma^j_{\mu\nu},
\end{aligned}\end{eqnarray}
where $M^{\beta,lm}_j$ are the matrices given in the orbital weight table. 
This can be seen by noting that
\begin{eqnarray}\begin{aligned}
\sum_{\mu,\nu}\rho^{\beta,lm}_{\mu\nu} \sigma^j_{\nu\mu}=\frac{1}{2}\text{tr}(\sum_{k=0}^3M^{\beta,lm}_k\sigma^k\sigma^{j}) =M^{\beta,lm}_j ,
\end{aligned}\end{eqnarray}
which reproduces Eq.~\eqref{eq:matrix_elemnt_3}. 
As a result, the spin-resolved density matrix $\rho^{\beta,lm}$ is completely determined by the given orbital weights $M^{\beta,lm}_j$. 
The sRDM is then given by tracing out the ion and orbital degrees of freedom
\begin{eqnarray}\begin{aligned}
\label{eq:def_spinRDM}
\rho_{\text{sRDM}} \equiv \sum_{\beta,lm}\rho^{\beta,lm}.
\end{aligned}\end{eqnarray}
We see then that we can write the sRDM as 
\begin{equation}
\rho_{\text{sRDM}} = \frac{1}{2}M_0\left(\sigma_0 + \mathbf{n}\cdot\vec{\sigma}\right),\label{eq:sRDM_from_dft}
\end{equation}
where we have introduced
\begin{align}
M_0 &= \sum_{\beta,lm} M_0^{\beta, lm}, \label{eq:numerical_sRDM_1}\\
n_x &= \frac{1}{M_0}\sum_{\beta,lm} M_1^{\beta, lm}, \\
n_y &= \frac{1}{M_0}\sum_{\beta,lm} M_2^{\beta, lm}, \\
n_z &= \frac{1}{M_0}\sum_{\beta,lm} M_3^{\beta, lm}.\label{eq:numerical_sRDM_2}
\end{align}
Eqs.~\eqref{eq:numerical_sRDM_1}--\eqref{eq:numerical_sRDM_2} are the formulas used to obtain the spin reduced density from DFT for the $d_{xy}, d_{yz}$ and $d_{zx}$ orbitals, as shown in the main text. To do so, we restrict the sums over $l,m$ to range only over the $d$-orbitals of interest. By factoring out $M_0$ in Eq.~\eqref{eq:sRDM_from_dft}, we account for the fact that the $d$-orbitals do not fully exhaust the trace of the density matrix.
The sRDM for other orbitals are shown in Fig.~\ref{fig:comparison_app}, which will be discussed in the next subsection. 

\subsection{$\alpha$-averaged sRDMs for sRDMs at the $R$-point for SG 198}
\label{sec:alpha-averaged}
In Sec.~\ref{sec:198_R}, we have shown that the most general ${\bf k\cdot p}$ Hamiltonian for the sixfold degeneracy near the $R$-point reads
\begin{eqnarray}
\begin{aligned}
    H_R({\bf \delta  k},b,\varphi, v_f)&=\hbar v_f\left(
    \begin{array}{cc}
    H^{(3)}(\delta\mathbf{k},\varphi) & bH^{(3)}(\delta\mathbf{k},0) \\
    b^*H^{(3)}(\delta\mathbf{k},0) & -H^{(3)}(\delta\mathbf{k},-\varphi)
    \end{array}
    \right), \\
    H^{(3)}(\mathbf{\delta k},\varphi)&=\left(\begin{array}{ccc}
    0 & e^{i\varphi}\delta k_z & e^{-i\varphi}\delta k_y \\
    e^{-i\varphi}\delta k_z & 0 & e^{i\varphi}\delta k_x \\
    e^{i\varphi}\delta k_y & e^{-i\varphi}\delta k_x & 0
    \end{array}\right),
\end{aligned}
\end{eqnarray}
where $v_f$ is the Fermi velocity, and $b, \varphi$ are free parameters not determined by symmetry.
As shown in Ref.~\cite{flicker2018chiral}, $H_R$ commutes with the operator
\begin{align}
A = \begin{pmatrix}
\cos\varphi & b\\
b^* & -\cos\varphi
\end{pmatrix}\otimes\mathbb{I}_3,
\end{align}
such that with the eigenstates of $A$, we can block-diagonalize the Hamiltonian as
\begin{widetext}
\begin{align}
\label{eq:six_fold_H_block}
H_R({\bf k},b,\varphi, v_f) = \sqrt{1+|b|^2}\hbar v_f\begin{pmatrix}
H^{(3)}(\frac{\pi}{2}-\delta\varphi,{\bf \delta k}) & 0 \\
0 & H^{(3)}(\frac{\pi}{2}+\delta\varphi,{\bf \delta k})
\end{pmatrix},
\end{align}
where $\delta\varphi=\tan^{-1}\left(\sqrt{\cos^2\varphi+|b|^2}/\sin\varphi\right)$. 
For $\delta\mathbf{k}$ oriented along the $(001)$ or $(011)$ directions, the eigenvalues of $H_R$ read
$
    \{0, \pm \hbar v_f\sqrt{1+|b|^2}|{\bf \delta k|}\}
$
which are doubly degenerate. 
Here $|{\bf \delta k|}$ is the norm of the momentum along the paths. 
For the $(111)$ direction, however, the eigenvalues read
\begin{align}
    \left\{\pm2\sin\delta\varphi \frac{|{\bf \delta k}|}{\sqrt{3}}, \pm(\sqrt{3}\cos\delta\varphi+\sin\delta\varphi)\frac{|{\bf \delta k}|}{\sqrt{3}}, \pm(\sqrt{3}\cos\delta\varphi-\sin\delta\varphi)\frac{|{\bf \delta k}|}{\sqrt{3}}\right\},
\end{align}
\end{widetext}
in units of $\hbar v_f\sqrt{1+|b|^2}$. These depend on $\delta\varphi$ explicitly. 
In order for the bands to be doubly degenerate, we require either $\tan\delta\varphi=0$ or $\tan\delta\varphi\pm\sqrt{3}$. 
In the first case, this results in $b=0, \varphi=\pm\pi/2$ which is identical to the case considered in Sec.~\ref{sec:198_R} and hence will not be considered here. 
Instead, we are interested in the second case, which results in the relation 
\begin{align}
\label{eq:relation_b}
    |b| = \sqrt{3\cos^2\varphi-\sin^2\varphi},
\end{align}
and the doubly degenerate spectrum
\begin{align}
\label{eq:double_spectrum}
    \{0, \pm 2\hbar v_f\cos\varphi|{\bf \delta k}|\},
\end{align}
for all three paths.
Eq.~\eqref{eq:double_spectrum} serves as a first order approximation for the DFT band structures shown in Fig.~\ref{fig:comparison_v2}(a). We will now determine the parameters.

\begin{figure}[H]
\includegraphics[width=\columnwidth]{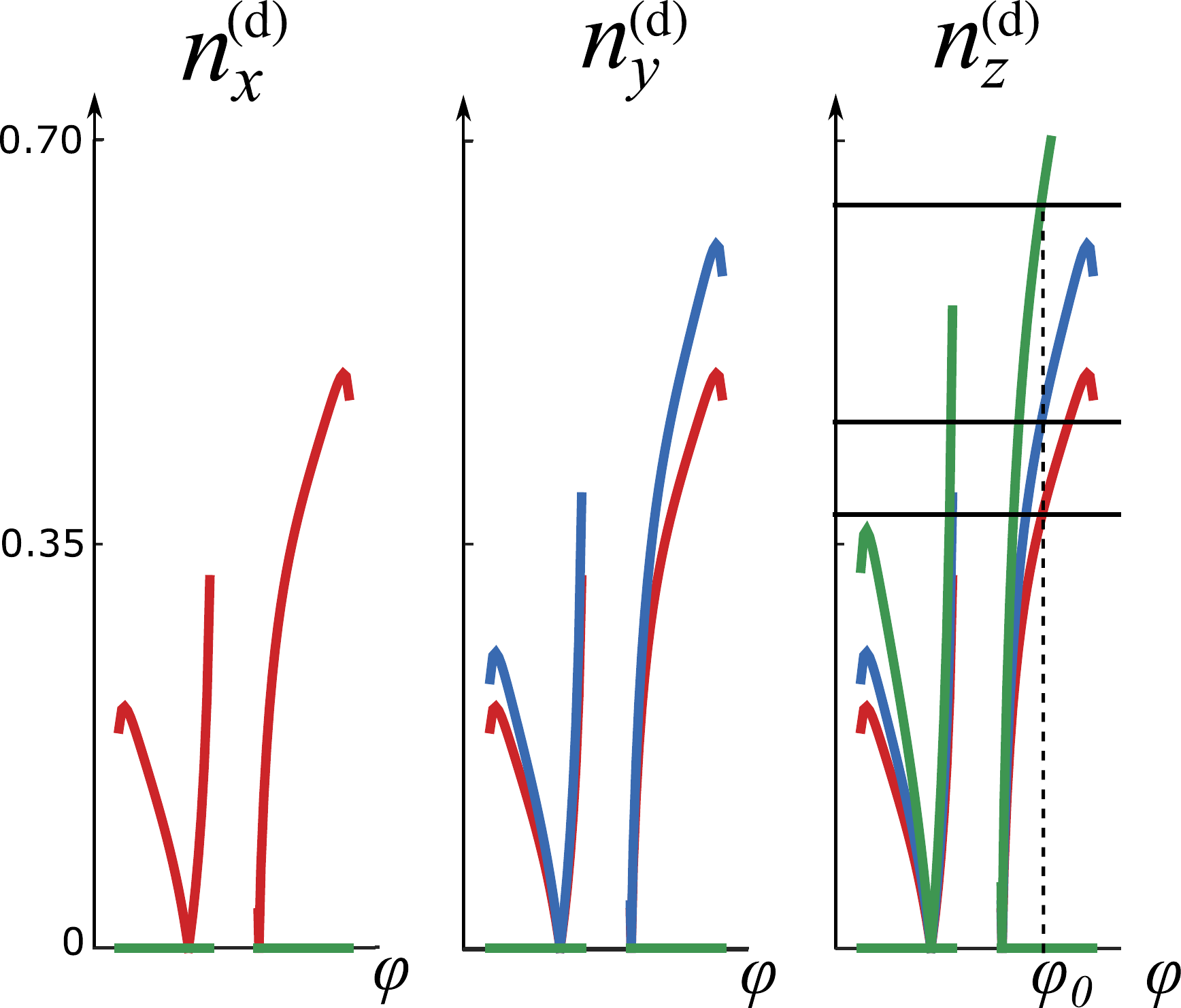}
\caption{The absolute value of the $\alpha$-averaged sRDMs for the $(d_{xy}, d_{yz}, d_{zx})$ subspace for the sixfold degeneracy at the $R$-point in PtGa. 
The three panels correspond to the $x,y$ and $z$-components of the sRDMs, and the red, blue and green solid lines correspond to the $\alpha$-averaged sRDMs for the paths $(111)$, $(011)$ and $(001)$, as a function of $\varphi$, respectively. 
The three black solid lines in the right panel correspond to the $z$-component of the sRDMs along the three paths. 
They intersect the colored solid lines at the same value of $\varphi_0=1.4280$, as indicated by the vertical dashed line. 
}
\label{fig:avg_v2}
\end{figure}

We proceed to first determine $\varphi$ by calculating the sRDMs for the sixfold degeneracy with the method developed in Sec.~\ref{sec:198_R}. 
For each value of $\varphi\in[-\pi, \pi)$, we determine $b$ from Eq.~\eqref{eq:relation_b} (it turns out that real values of $b$ is sufficient for our purposes), followed by diagonalizing $H_R(\delta{\bf k},b,\varphi, v_f)$. 
The eigenstates can be labeled as $|\alpha=\pm, \delta{\bf k}m\rangle$ with eigenvalues $E(\alpha=\pm, \delta{\bf k}m)=2m\hbar v_f\cos\varphi|{\bf \delta k}|$ for $m=0,\pm1$. 
Since the eigenstates are doubly degenerate and do not depend on $\alpha$, we expect that the measured sRDM for certain value of $m$ will be the average over $\alpha$. 
In Fig.~\ref{fig:avg_v2}, we show the absolute value of the $\alpha$-averaged sRDMs for the $(d_{xy}+d_{yz}+d_{zx})$ subspace. 
We note that from Eq.~\eqref{eq:relation_b}, $b$ is only well-defined for $\varphi\in[-\frac{2\pi}{3}, -\frac{\pi}{3}]\cup[\frac{\pi}{3}, \frac{2\pi}{3}]$, and hence the sRDMs are piecewise functions of $\varphi$. 
The different branches of the sRDMs for a given path, say $(111)$, correspond to different magnetic quantum $m$. 
In order to compare to the DFT results, we extract the $z$-components for each path, and plot them as black solid lines on top of the theory predictions in the right panel of Fig.~\ref{fig:avg_v2}. 
We see that if we choose $\varphi_0=1.4280$ to match the value of $n_z^{(d)}$ along the $(001)$ direction from DFT, we immediatley get agreement with our theoretical predictions for $n_z^{(d)}$ along the $(011)$ and $(111)$ direction. 
Alternatively, the same value of $\varphi_0$ can be determined if we compare, say, the $x$-component of the sRDMs.
With the value of $\varphi_0$ determined, it is straightforward to determine $v_f$ by fitting the band structures in Eq.~\eqref{eq:double_spectrum} to that from the DFT result. 
We find that $hv_f/a=2.4$ meV, and the resulting band structures are shown in Fig.~\ref{fig:comparison_v2} in the main text. 

\begin{widetext}

\begin{figure}[H]
\includegraphics[width=\columnwidth]{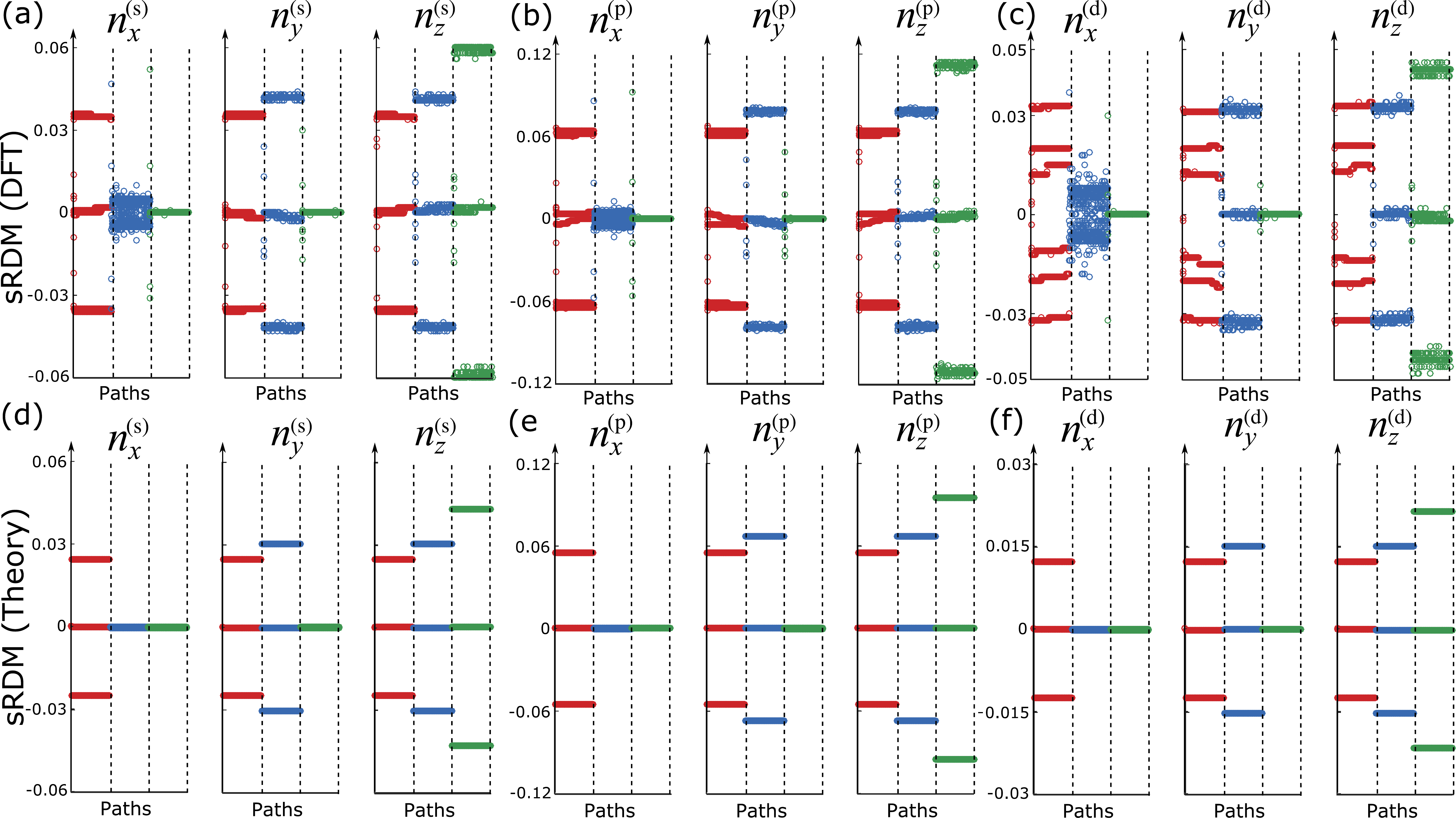}
\caption{Comparison of predicted spin-momentum locking with results from ab-initio DFT calculations. 
The paths are the same as the ones shown in Fig.~\ref{fig:comparison_v2}.
(a-c) DFT results for $s$, $p$ and $d_{x^2-y^2}, d_{z^2}$ orbitals. 
Note that the label $n_i^{(d)}$ refers to the $d_{x^2-y^2}, d_{z^2}$ orbitals, instead of $d_{xy}, d_{yz}, d_{zx}$ as in Figs.~\ref{fig:comparison_v2}--\ref{fig:avg_v2}.
(d-f) Theory results for the same orbitals. 
The results for the $s$, $p$ and $(d_{x^2-y^2},d_{z^2})$ orbitals have been divided by $3.40$, $6.81$ and $15.31$ according to their relative orbital weights compared to the $d_{xy}+d_{yz}+d_{zx}$ orbital.
}
\label{fig:comparison_app}
\end{figure}

\end{widetext}

In Fig.~\ref{fig:comparison_app}(a-c), we show the DFT sRDMs for the $s$-orbital, $p$-orbitals and the $d_{x^2-y^2},d_{z^2}$ orbitals respectively. 
The corresponding theory results, namely the $\alpha$-averaged sRDMs, are shown in Fig.~\ref{fig:comparison_app}(d-f).
For the $p$-orbital (with orbital weight around 11\%), similar to the $d_{xy},d_{yz}$ and $d_{zx}$ orbitals discussed in the main text, we find the $p_x,p_y$ and $p_z$ orbitals share the same weights, and hence the main contribution comes from the one-dimensional representation $p_x+p_y+p_z$.
The theory result has been divided by a factor of $6.81$, which is the ratio between the orbital weights of the $d_{xy}+d_{yz}+d_{zx}$ orbital and the $p_x+p_y+p_z$ orbital.
Similarly, for the $s$ and $(d_{x^2-y^2}, d_{z^2})$ orbitals, we have divided the theory result by $3.40$ and $15.31$ in order to faithfully compare to the DFT results. 
From Fig.~\ref{fig:comparison_app}, we see that due to the relatively large orbital weights, the results for the $s$ and $p$ orbitals agree quantitatively. 
The theoretical predictions for the $(d_{x^2-y^2}, d_{z^2})$ orbitals are significantly different from the DFT result, due to their relatively low orbital weights (only $\sim 3.2$\%). We expect that this does not affect our quantitative understanding for the overall sRDM, when compared to experiments.

\bibliography{refs}
\end{document}